\documentclass[letterpaper]{article} 
\makeatletter
\def\input@path{{AnonymousSubmission/LaTeX/}}
\makeatother
\usepackage{aaai2026}  
\usepackage{times}  
\usepackage{helvet}  
\usepackage{courier}  
\usepackage[hyphens]{url}  
\usepackage{graphicx} 

\graphicspath{{AnonymousSubmission/fig/}{AnonymousSubmission/LaTeX/}}

\usepackage{natbib}  
\usepackage{caption} 
\usepackage{algorithm}
\usepackage{algorithmic}
\usepackage{shortcuts}
\usepackage{booktabs}
\usepackage{tikz}
\usepackage{nameref} 

\usetikzlibrary{
  arrows.meta,
  positioning,
  calc,
  decorations.pathreplacing,
  shapes.misc,  
  matrix 
}
\usepackage{mathtools}
\usepackage{mdframed}
\usepackage{amssymb}
\usepackage{amsmath}
\usepackage{amsthm}
\usepackage{thmtools}
\usepackage{subcaption}
\usepackage[table]{xcolor}
\definecolor{lights}{HTML}{D8E9FF} 
\definecolor{lightdf}{HTML}{fec44f} 
\definecolor{gree}{HTML}{a8ddb5}
\definecolor{greeey}{HTML}{bdbdbd}

\usepackage[most]{tcolorbox}
\newtcolorbox{mybox}[2][]{%
  attach boxed title to top center
               = {yshift=-8pt},
  colback      = black,
  colframe     = gree,
  fonttitle    = \bfseries,
  colbacktitle = gree,
  coltitle=black,
  title        = #2,#1,
  enhanced,
}
\newtcolorbox{myboxd}[2][]{%
  attach boxed title to top center
               = {yshift=-8pt},
  colback      = black,
  colframe     = lights,
  fonttitle    = \bfseries,
  colbacktitle = lights,
  coltitle=black,
  title        = #2,#1,
  enhanced,
}
\definecolor{vscodebg}{HTML}{1F1F1F}
\newtcolorbox{myboxf}[2][]{%
  attach boxed title to top center
               = {xshift=-16pt,yshift=-8pt},
  colback      = vscodebg,
  colframe     = greeey,
  fontupper=\color{white},
  fonttitle    = \bfseries,
  colbacktitle = greeey,
  coltitle=black,
  title        = #2,#1,
  enhanced,
}

\declaretheoremstyle[
headfont=\normalfont\bfseries,
notefont=\mdseries, notebraces={(}{)},
bodyfont=\normalfont,
postheadspace=0.25em,
spaceabove=1pt,
mdframed={
  skipabove=8pt,
  skipbelow=6pt,
  hidealllines=true,
  backgroundcolor={lights},
  innerleftmargin=0pt,
  innerrightmargin=0pt}
]{shaded}
\declaretheorem[style=shaded]{definition}

\usepackage{xcolor}

\newcommand\gs[1]{\textcolor{blue}{[Grant: #1]}}
\newcommand\zx[1]{\textcolor{orange}{[Zexuan: #1]}}
\makeatletter
\newcommand*\bigcdot{\mathpalette\bigcdot@{.5}}
\newcommand*\bigcdot@[2]{\mathbin{\vcenter{\hbox{\scalebox{#2}{$\m@th#1\bullet$}}}}}
\makeatother

\usepackage{newfloat}
\usepackage{listings}
\DeclareCaptionStyle{ruled}{labelfont=normalfont,labelsep=colon,strut=off} 
\floatstyle{ruled}
\newfloat{listing}{tb}{lst}{}
\floatname{listing}{Listing}
\nocopyright

\title{ \ourmethode{}: Analyzing a human-LLM system for annotating social media data with the concept of climate change mitigation pessimism}

\author{
    Zexuan Li\textsuperscript{\rm 1},
    Derek Van Berkel\textsuperscript{\rm 2},
    Ariel Hasell\textsuperscript{\rm 3},
    Grant Schoenebeck\textsuperscript{\rm 1},
    John Barry Ryan\textsuperscript{\rm 3},
    Sabina Tomkins\textsuperscript{\rm 1}
}

\affiliations{
    \textsuperscript{\rm 1}School of Information, University of Michigan\\
    \textsuperscript{\rm 2}School for Environment and Sustainability, University of Michigan\\
    \textsuperscript{\rm 3}Department of Communication and Media, University of Michigan\\
}

\usepackage{bibentry}

\begin{document}

\maketitle

\begin{abstract}
Large language models (LLMs) are increasingly being integrated into research workflows.
However, LLMs have been shown to struggle with difficult and nuanced concepts such as those found in computational social science (CSS) research.
Within the CSS community, there has been a call for new systems to  be developed which center humans in LLM-supported scientific workflows. 
We develop  \ourmethode{}, a human-centered system for inspecting and improving LLM annotations, a process we refer to as \align{} for a target concept. 
\ourmethode{} is developed with both computational and social scientists to reflect existing workflows for data annotation. It includes a range of information features for users to interrogate the quality and reliability of LLM annotations.  
We evaluate our system in two settings. In the first, we assume a researcher may not have access to ground truth data and that users of \ourmethode{} have limited prior knowledge of the concept they would like an LLM to annotate. That is, they may be conducting \spec{} and \align{} simultaneously. 
In the second setting, we assume access to ground truth labels and that the concept is specified for a given annotation task, here the task of \align{} is more straightforward. 
We find that in both settings users can improve the quality of LLM annotations with \ourmethode{} and that their final annotations far surpass those created without human intervention. For example when we evaluate with ground truth labels we see an absolute improvement of 0.15 in F-Measure and 0.23 in accuracy over a fully automated state-of-the-art method for prompt refinement. 

\end{abstract}

\section{Introduction}
 While some have 
 proposed that large language models (LLMs) can serve as zero-shot annotators in natural language processing (NLP) settings \citep{chae2023large,zhang-etal-2025-efficient}, there is less evidence that they can be useful for complex annotations such as those required by the social sciences \citep{guo2024evaluating,karimi-etal-2021-aeda-easier,bucher2024fine,felkner2024gpt,nasution2024chatgpt,pangakis2025keeping,zamfirescu2023johnny}. 
Instead, there have been calls for human-centered workflows for annotation \citep{shin2024understanding,li2023coannotating,ranade2025using,ma2025should,he2024if}.
Our work responds to these calls with a system, \ourmethode{}, which is designed for computational social science (CSS) research settings.
The goal of \ourmethode{} 
is to support researchers in grounding LLMs in their concept definition, both when that concept definition is still undergoing specification and when it is well-specified. 

Consider a researcher with an idea of a concept they would like to mine in social media data, such as climate pessimism. Typically, they may explore the data and refine the concept through examples. Next, they may describe this concept to a group of research assistants and iteratively develop a codebook to specify how the concept should be coded. Here, they may still refine the concept through feedback from the labelers. We refer to the first process as \spec, where the concept is defined in increasingly concrete terms. 
We refer to the second process as grounding, where the goal is to align labelers with the new specification.    
We bring together social scientists from diverse fields and computational scientists to design information features for interrogating LLM outputs which support \align{} to both evolving and existing specifications.
These features are designed to mimic key aspects of the social science annotation pipeline, such as evaluating label consistency and reliability, and providing natural language explanations.



We  evaluate our system in a setting with high social impact and where we expect LLMs to struggle: the task of detecting pessimism with respect to the ability to mitigate the effects of climate change (\climate). 
The  abundance of negative political content on social media platforms may
 cause users  to feel overwhelmed, anxious, and fatigued with news and politics,  leading to cynicism and lower levels of political efficacy \citep{lane2025worn,hasell2025social,song2021social}.
 Yet, it is an open question how this relates to attitudes towards climate change. 
Like others \citep{galdeman2025mapping,pearce2019social,treen2022discussion,hautea2021showing}, we utilize social media data  as a key source for insights into public attitudes about climate change. 

We design two studies to evaluate \ourmethode. In the first, we assume that the target concept is not yet specified. This represents the case where specification and grounding may occur simultaneously, such as when a researcher iterates with labelers to develop a codebook. 
In the second, we assume that the concept has been specified. Here, \subs{} have access to a  limited amount of ground-truth labels which they can use to evaluate the LLM. 
In both cases we evaluate the ability of \ourmethode{} to support \align, asking whether \subs{} can improve the ability of LLMs to detect \climate{} through interaction with our system. 
We recruit 60 participants from a graduate program on the environment and 
sustainability to use our system to detect \climate{} in TikTok posts.


 We present \ourmethode{} as a CSS contribution and evaluate it on a carefully selected case study. 
 We also introduce the problem of detecting \climate, an important attitude that integrates sustainability, communication, and political science research. We find that while \ourmethode{} is effective for improving LLM outputs, there is still room to improve on the task of detecting \climate.

\section{Why climate change mitigation pessimism?}
Climate change mitigation pessimism was selected as the focus of this study for several reasons. First, it reflects a growing concern among climate researchers and communicators \citep{davidson2023}.
Additionally, climate change mitigation pessimism aligns with the academic and personal interests of a broad participant pool. 
We are especially interested in understanding the experiences of non-technical users, and it is important to find a concept of relevance to a wide group of sustainability researchers.

Second, the concept presents methodological challenges that make it particularly well-suited for evaluating information-supported annotation workflows for social media data. Climate change mitigation pessimism is subjective, emotionally engaged, and often embedded in subtle language or tone. It may be expressed implicitly — through sarcasm, humor, or despair — rather than through clear propositional statements. 
This makes it difficult to annotate using rule-based heuristics or simple keyword matching. Because interpretations of pessimism can vary across disciplinary backgrounds and personal experiences, the annotation task involves active sensemaking and specification \citep{pirolli2005, dork2011}.

Our research team consists of three social scientists 
(two with training in political science and communication, one in sustainability), 
two computational social scientists, and one graduate student in human-centered computing. 
Together the team has expertise in the target concept of climate change mitigation pessimism,
both from the perspective of climate adaptation and the ethical use of social media in sustainability science, 
and the perspective of how the public responds to science information in social media environments.  


We define climate change mitigation pessimism  as expressions of the belief that humans cannot (or will not) meaningfully reduce the causes or impacts of climate change. The tone is typically hopeless or resigned and may imply giving up on mitigation efforts. Social media users may express this belief in posts about the failure of government to take meaningful action against climate change, or in statements such as ``it is too late'' to do anything about this problem. We operationalize this as shown in \defnref{def1}.

\begin{definition}[\textbf{Climate Change Mitigation Pessimism}]
 Pessimism with regard to climate change mitigation (the ability to take action to reduce the event
of and effects of climate change). \label{def1}
\end{definition}

In particular, we are interested in how this concept is discussed by the general public on social media platforms, which are becoming increasingly important sources of news and information.  
To ground the study, we focused on TikTok, a popular and unique platform, which has become an important source of news and information for young adults; almost 40\% of them rely on it regularly\footnote{\url{https://www.pewresearch.org/short-reads/2024/09/17/more-americans-regularly-get-news-on-tiktok-especially-young-adults/}}.

\commentout{
Throughout the process the team discussed annotation workflows they have experienced before, as well as what would be required to judge LLM labels. These discussions led to the following goals.  
With these goals in mind, we iteratively developed \ourmethode. 

\begin{center}
\begin{myboxd}[colback = white, width = 8cm]{\ourmethode{} - Goals}

 (1) \ourmethode{} should enable \subs{} to observe how an LLM would react to their instructions and to refine their instructions \textbf{iteratively} in response to these observations.

  (2) \ourmethode{} should provide \subs{} with information features with which they can judge the \textbf{reliability} of LLM outputs.

 (3) \ourmethode{} should provide \subs{}  \textbf{explanations} in the way that mirrors the explanations they would receive from a research mentor, such as a social science expert.

 (4) \ourmethode{} should provide \subs{} with guidance about how social scientists would behave (\textbf{expert scaffolding}), to reduce and counter potential overreliance on LLM outputs and improve understanding of a complex and \textbf{subjective} concept. 

\end{myboxd}
\end{center}
}

\section{\ourmethode{} Overview}
\begin{figure*}[t]
  \centering
  \includegraphics[width=.6\textwidth]{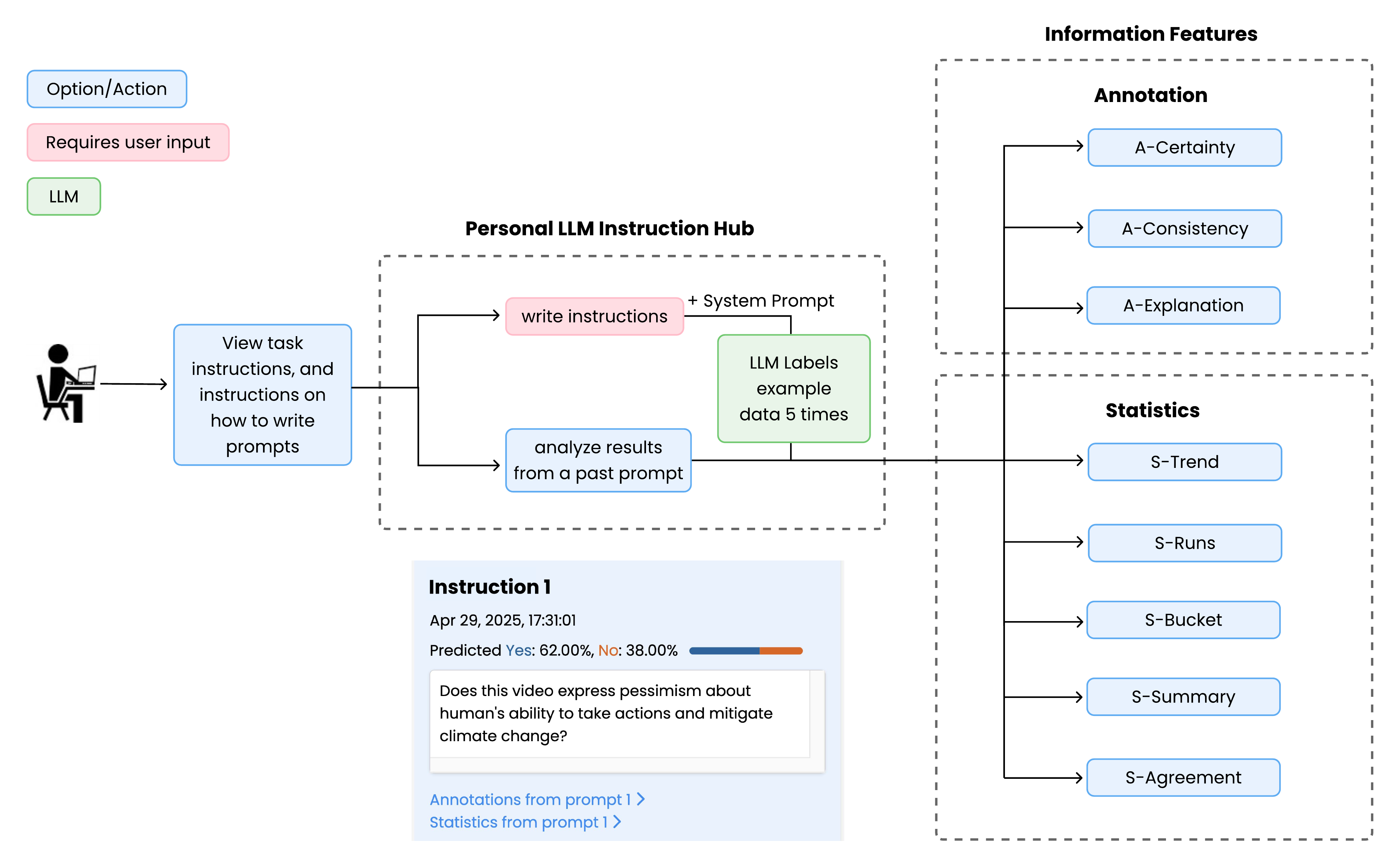}
  \caption{\ourmethode{} system workflow. Participants iteratively write instructions, revisiting the instruction hub and information features many times. }
  \label{fig:overview}
\end{figure*}

\figref{fig:overview} shows how \subs{} interact with \ourmethode.
First, \subs{} view task instructions. These instructions include optional links they can follow to read more detailed instructions about guiding LLMs, compiled from academic websites on prompt design. 
Next, \subs{} are directed to their personal LLM Instruction hub (\prm{}). Here, they can write instructions and view information about past instructions they've written. 
After instructions are written and submitted, they are appended to a system prompt (see Appendix~\nameref{app:systemPrompt}).
Whenever a participant submits a prompt, this new prompt (which includes the participant's instructions and the system prompt) is sent to one LLM five times\footnote{The prompt is sent to the LLM five times to create reliability and consistency measures.}, along with a validation set of TikTok videos which the LLM is asked to annotate. 

The LLM then determines if each of the videos contains the concept of \climate{} (where it is prompted to answer ``Yes'' if the video contains the concept and ``No'' if it does not), provides an estimate of how certain it is about whether the video contains the concept, and generates an explanation of its decision. 
Once the annotations of the videos have been completed (approximately 30-120 seconds), there are two pages \subs{} can navigate to from \prm. 

\begin{figure*}[t]
    \centering
    \fcolorbox{gray!40}{white}{
    \includegraphics[width=.8\textwidth]{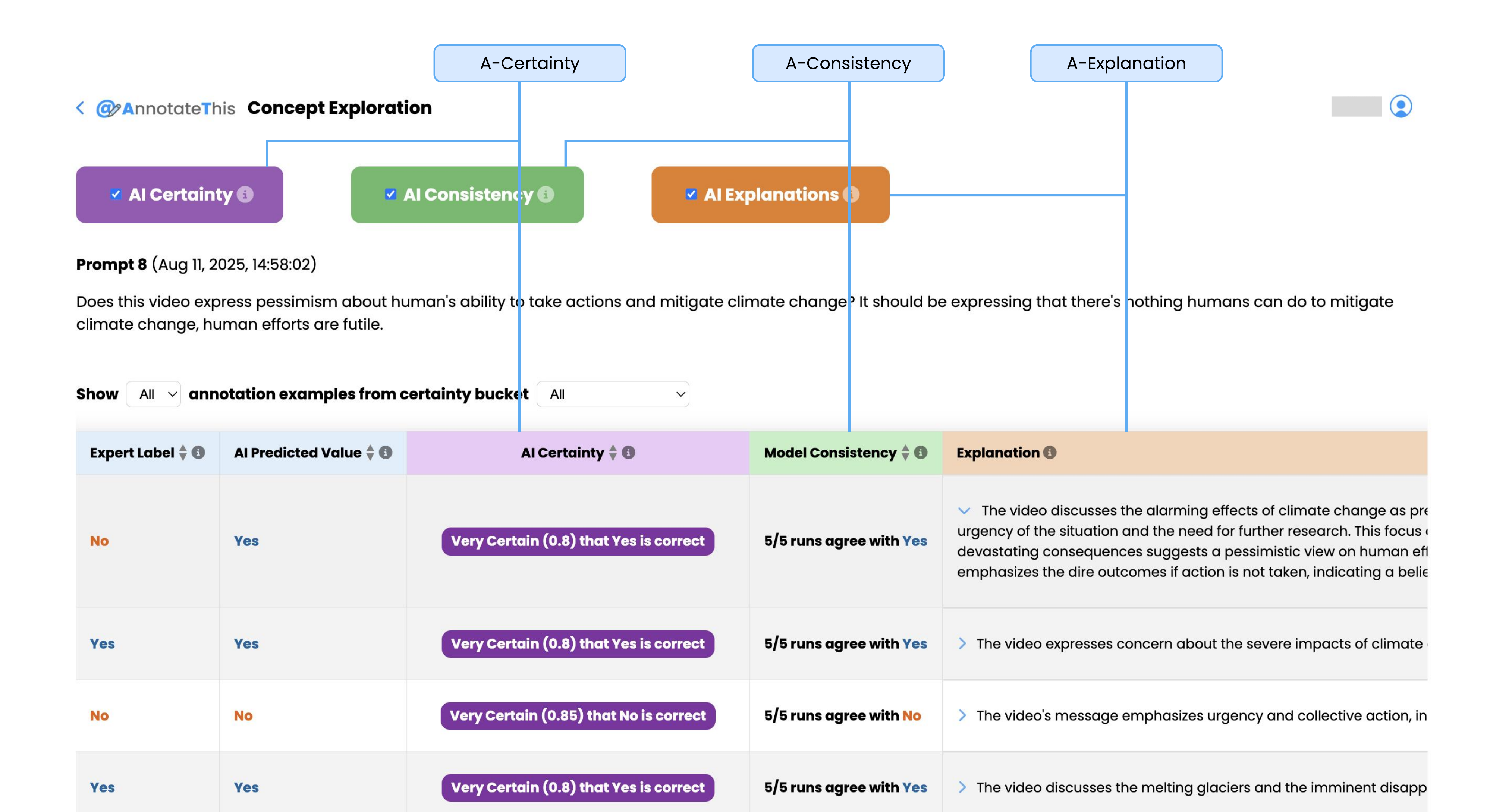}
    }
    \caption{Screenshot of \ann{} view with information features, the scrollable table contains video URLs and transcripts to the left. Here the expert label refers to the ground truth produced by the research team. This label was only shown in \stwo, in \sone, we assume ground truth labels do not exist. 
    }
    \label{fig:ann-screenshot}
\end{figure*}

The first page is the annotations results and explanations page (\ann{} shown in \figref{fig:ann-screenshot}). This page displays a table where each row corresponds to a single TikTok video (A-Table). The table displays both non-optional columns (static features) and optional columns which \subs{} can choose to see (information features).
The static features are: a table index, a video url, a video transcript,  an expert label produced by the study team (which is only shown in \stwo), and the AI label
(a screenshot of the table with the static features is shown in \figref{fig:screenshot_anntable} in Appendix~\nameref{app:system_more}). 
For each video the table can additionally display three information features which \subs{} can reveal by selecting a box as shown in \figref{fig:overview}:
\begin{description}

    \item[A-Certainty] a certainty estimate about the AI label\footnote{For each instance, the system prompt asks the LLM to provide a numerical estimate of its certainty on its answer on a scale of 0 to 1. This is bucketed into five qualitative ranges, from ``Very Uncertain'' to ``Very Certain'', derived  from \citet{tian_just_2023}.}, 
    \item[A-Consistency] an estimate of model consistency,
    \item[A-Explanation] an explanation for the AI label which is generated by the LLM.
\end{description}

These information features were designed to provide information about label quality deemed important by social scientists, so that (1) \subs{} have direct knowledge of the data itself through the video links and transcripts, 
(2) \subs{} have two measures of the reliability of the generated label - whether one of the five LLM runs is confident in the label and whether the LLM is consistent in the label it generates across 5 runs, and 
(3) \subs{} have direct exposure to the ``thought process'' of the LLM through natural language explanations. 

The second page \subs{} can navigate to is the statistics page (\sta)\footnote{Screenshots of each information feature on the \sta{} page are shown in the Appendix~\nameref{app:infofeatures}.} 
This page provides information that allows \subs{} to consider how the LLM is behaving under different instructions. Here they see: 


\begin{description}
    \item[S-Trend] this is designed to show \subs{} how uncertain the LLM is across different instructions, it displays a line graph of uncertainty outputs for each set of instructions,
    \item[S-Runs] this is designed to show how the label distribution varies across  model runs, it displays a bar chart of the number of instances with each label (``Yes'' or ``No'') across the 5 model runs for each instruction,
    \item[S-Buckets] this is designed to show how the LLM's self-reported certainty varies across labels, it displays a bar chart with the count of predictions for each certainty bucket broken down by label (``Yes'' vs. ``No''),
    \item[S-Summary] this is designed to surface themes in the LLM generated explanations, it displays a summary of all explanations produced from a given instruction,
    \item[S-Agreement] this is designed to test if there is a relationship between the LLM's self-reported certainty and the consistency of the model across the 5 runs, it displays a scatterplot of model agreement and certainty and additional information.
\end{description}
\vspace{-2mm}
\begin{table}[h]
  \centering
  \caption{Panel and widget tags used throughout the paper.}
  \label{tab:codes}
  \begin{tabular}{p{2.2cm}p{4.8cm}}
    \toprule
    \multicolumn{2}{l}{\textbf{Personal LLM-I hub (\prm{})}} \\
    \midrule
    — & (no widgets analysed) \\[6pt]

    \multicolumn{2}{l}{\textbf{Annotation page (\ann{})}} \\
    \midrule
    \anncert{} & LLM certainty for each instance\\
    \anncons{} & Consistency across 5 runs\\
    \annexpl{} & Natural-language explanation\\
    \anntabl{} & Sortable annotation table\\

    \multicolumn{2}{l}{\textbf{Statistics dashboard (\sta{})}} \\
    \midrule
    \statren{}  & Uncertainty trend line     \\
    \staruns{} & Label-change bar (5 runs)     \\
    \stabuck{}  & Uncertainty for all labels  \\
    \staexsm{}  & Explanation summary \\
    \staagrm{}  & Certainty vs agreement scatter  \\
    \bottomrule
  \end{tabular}
\end{table}
\vspace{-3mm}
\paragraph{Implementation details}
The front end is developed with SvelteKit and deployed via Fly.io. The front end communicates with a backend hosted on Amazon AWS RDS. The database records user credentials (with bcrypt encryption), submitted prompts, video metadata, model outputs, and user interaction logs (e.g., button clicks). Model inference is powered by OpenAI's GPT-4o-mini (release: 2024-07-18) accessed through the OpenAI API \citep{openai_gpt4omini,openai_api_docs}.

\commentout{

\subsection{\ourmethode{} System Overview}
\label{sec:soe}
In addition to the information features provided in the base version (listed in \tabref{tab:codes}), we provide three forms of expert scaffolding: 
The first is a rewording of the term \climate. 
In \sone, \subs{} are told that they are training an LLM to detect ``climate pessimism'' and given the definition above (\defnref{def1}).

In \stwo, we continue to use \defnref{def1}, however rather than referring to climate pessimism in all training materials and in \ourmethode, we refer to \climate. Incorporating the word mitigation directly into the target concept was intended to reinforce that \subs{} are not only being asked to think about attitudes towards climate change, but attitudes towards humans' ability to mitigate its consequences. 

The second addition is a training quiz. Participants are shown five videos and asked to determine if the video contains \climate. 
They are then provided feedback about how social science experts would label the video along with the experts' rationale. 

Finally, we add one additional column to A-Table (Expert Label). 
Before the column showing the LLM's label, we include a column showing a label (``Yes'' or ``No'') produced through  consensus of the entire research team. This is meant to provide an alternative anchor to the outputs produced by the LLM.  

}

\section{Empirical Evaluation}
We evaluate \ourmethode{} in two settings. 
First, we consider the case where a concept is not yet specified in \sone. Here, we expect that 
\subs{} will use \ourmethode{} for \spec{}  while iterating on \align. 
This design mimics the process where a researcher develops a codebook through iteration with different labelers. In this case, LLMs may serve as a stand-in for human annotators in the process of codebook design. 
In \stwo, we consider the case where \subs{} have an existing specification and are using \ourmethode{} to ground the LLM in that specification. 

\subsection{Social media data collection} 
We constructed a dataset of TikTok videos relevant to climate change mitigation pessimism through a two-stage scraping and refinement process. The team initially curated a hierarchical list of climate-related terms, including umbrella concepts such as \textit{climate change}, \textit{extreme weather}, and \textit{biodiversity loss}, and more specific terms under each umbrella e.g. \textit{carbon}, \textit{global warming}, and \textit{greenhouse gases} would be specific terms under \textit{climate change}. 
Using these terms, we collected a preliminary set of TikTok videos using the TikTok Research API. We  analyzed these videos for frequent keywords, n-grams, and keyword co-occurrence patterns in order to select a final set of key words for collecting the dataset for the studies which is shown in Appendix~\nameref{app:keywords}. 
\commentout{Using these terms, we collected 21{,}771 TikTok videos using the TikTok Research API  and conducted an exploratory analysis of frequent keywords, n-grams, and keyword co-occurrence patterns. We also sampled representative videos for each keyword to aid expert refinement of the term list.

This process was repeated in a second round with 20{,}435 additional videos based on the revised term set. The final keyword list was selected after reviewing the initial set of videos. 
That is, the team then met together and analyzed the videos. In particular, we discussed how many videos were surfaced for each key term and whether these videos were actually relevant to climate change. }

Using this list, we collected a total of 62{,}513 TikTok videos published between January 1, 2024 and December 31, 2024.
To ensure content relevance, a random subset of 1{,}000 videos was annotated using GPT-4o-mini to determine the presence of three key climate-related themes: (1) effects or harms of climate change, (2) actions to mitigate climate change, and (3) scientific facts about climate change. We used the 185 videos which had at least one of these three themes  in the studies.
For example, a subset was given to the research team to determine ground truth labels, a subset was used for \subs{} to label (test set), a subset was used for LLMs to annotate (validation set), and a subset was used for the training quiz for \subs{} in \stwo.

\paragraph{Climate change mitigation pessimism ground truth}
Each ground truth label is the result of deliberation across the team. Each team members annotated 100 videos themselves, then we met and discussed each label for the videos which would be used in the studies until consensus was reached. 
In these 100 videos, we detected \climate{} in 18.7\% of posts and the agreement before deliberation was 74\%, with Fleiss Kappa of 0.429. Note that the low Kappa is influenced by the class imbalance of this label.


\subsection{Study Statistics}
In both studies we recruited graduate students in a program of study on sustainability and environmental science.
In both cases the sessions ran for no more than 2 hours and participants received a \$50 virtual credit card. 
Both studies were deemed exempt by an institutional IRB. 
After each study we asked \subs{} to fill out a survey. This included questions about their technical expertise (additional responses to this survey can be found in Appendix~\nameref{app:participants}). The majority  of \subs{} (55 out of 60) reported that they had no prior experience training LLMs to analyze text data. \sone{} was held in a computer lab in Spring 2025 and 27 \subs{} participated. 
\stwo{} was held on Zoom in Summer 2025 with 33 \subs.

\subsection{Study Flow}
Within the studies we use multiple datasets. At each iteration \subs{} can view LLM outputs on a validation dataset of 50 videos in \sone{} and 20 videos in \stwo. 
This validation set is distinct from the test set of 10 videos which they are asked to label themselves, both before and after the study. 
Participants followed a structured six-step workflow (full details in Appendix~\nameref{app:tasks}): 
\begin{enumerate}

    \item Consent form and study introduction.
    \item Training quiz on 6 practice videos (\stwo{} only).
    \item Label 10 TikTok videos for climate change mitigation pessimism.
    \item Iterate with \ourmethode: write instructions, view LLM predictions and information features, revise instructions. At each iteration the LLM labels the set of validation videos. 
    \item Re-annotate the same 10 videos as in step 3.
    \item Complete a brief post-task survey.
\end{enumerate}

\commentout{
In \sone, \subs{} were asked to  detect \climate{} in the same 10 videos before and after
using \ourmethode. 
Each participant would write   instructions for the LLM. 
These instructions were combined with the system prompt and an  LLM was asked to provide outputs for the validation set
5 times. These 5 outputs were aggregated in different ways according to the information feature. 
For example, the LLM label that \subs{} saw was the majority vote label across the 5 outputs. 

In \stwo, \subs{} were also asked to detect \climate{} in the same 10 videos before and after
using \ourmethode. 
In this case, the  LLM was asked to provide outputs for 20 videos
5 times.
}
In \stwo{} we reduced the size of the validation set from 50 to 20 as we saw that \subs{} rarely viewed 
more than  20 videos in \sone, and to improve the running time of the experiment. 
We provided ground truth labels for these 20 videos in \anntabl, as shown in \figref{fig:ann-screenshot} (Expert Label).

\subsection{Study 1}
In this initial study our goal was to understand if \ourmethode{} can be used to improve \align{} to \climate{} when \subs{} are new to the concept. Here, they don't have access to ground truth and may be conducting \spec{} while instructing the LLM. 
We frame this around the following research question, 
\textbf{RQ1:} Did \subs{} use \ourmethode{} to improve the LLM grounding in their interpretation of \climate{}? 

We evaluate this question in two ways. We assume that there is no canonical ground truth 
definition of the target concept. This represents a difficult case 
where concept specification is entangled with the task of grounding. 
However, this is common in research settings where ambiguous concepts become refined
through interaction with labelers. For example, a researcher may simplify a target concept in order to improve the ability to instruct labelers in how to identify it.  
We ask \subs{} to label videos 
according to their own interpretation of \defnref{def1}. 
They do this both before and after interacting with \ourmethode.
We can thus inspect whether their later LLM instructions are better able to 
produce LLM outputs which align with either their initial or final labels than 
their initial LLM instructions. This serves as a rudimentary check on the ability of \subs{} to use our system to learn how to ground LLMs in a target concept. 

Then, for each participant, indexed by $j\in J$, we generate a set of labels - that is the labels they assign to the videos, 
$y_j^{pre}$,  before encountering \ourmethode{} and a separate set of labels, $y_j^{post}$, which they assign after using \ourmethode{}.
For the first set of instructions they write, an LLM supplied with these instructions and the same set of videos 
generates labels  $y^0_j$.
Similarly, we  produce a set of LLM labels for each participant after $k$ iterations, $y^k_j$. 
To calculate the average change in accuracy between the final and initial instructions for \sone{} we compute the following: 

$$\frac{1}{|J|}\sum_{j\in J}(Accuracy(y_j^k,y_j^*)-Accuracy(y_j^0,y_j^*)),$$
where  $y_j^*$ denotes the ground truth. Here, we can consider either $y_j^{pre}$ or $y_j^{post}$ to be ground truth.

\begin{table*}[t]
\small
\caption{The performance metrics for \sone{}. 
We inspect how well the LLMs perform on their initial and final prompts against \subs{} labels before and after using \ourmethode{}. We bold all differences which
are statistically significant (p $<$ 0.05 according to Wilcoxon signed-rank tests).
} 
\label{tab:sone_performance}
\begin{tabular*}{\textwidth}{@{\extracolsep{\fill}}lcccccc}
\toprule
\multicolumn{7}{c}{RQ1: Did \subs{} use \ourmethode{} to improve the instructions they give an LLM?}\\
 \multicolumn{7}{c}{$\Delta=$ $\big($performance of  final prompt$\big)$ $-$ $\big($performance of initial prompt$\big)$}\\
 & &\textbf{$\Delta$Acc} & \textbf{$\Delta$Prec} & \textbf{$\Delta$Recall} & \textbf{$\Delta$F-Measure} & \textbf{$\Delta$ Avg Prompt Length}\\
\midrule
 &
 Ground truth $y_j^* = y_j^{pre}$ (initial labels)
 & \textbf{0.063} & \textbf{0.082} & \textbf{-0.095} & 0.019 & 395.6\\
 & 
 Ground truth $y_j^*=y_j^{post}$ (final labels)
 & \textbf{0.115} &  \textbf{0.112} & \textbf{-0.072} & \textbf{0.049} & 395.6\\


\midrule
\end{tabular*}
\end{table*}

In \tabref{tab:sone_performance} we see that 
there is a clear improvement in the instructions that \subs{} write when using \ourmethode. 
This is true whether we consider $y_j^{pre}$ or $y_j^{post}$ to be the ground truth.
We also see that \subs{} adapt their understanding of the concept over time, 
such that there is more improvement when we consider the final annotations. Whether this reflects a reliance on the LLM's outputs or not requires future study. For example, 
\subs{} may be learning from the LLM and exposure to the data, or they may be simply biased towards the LLM. 

The  length of \subs' instructions also increased substantially over time, with initial instructions averaging 237 characters and final instructions averaging 632 characters. This lengthening reflects more detailed instructions, richer definitions, and inclusion of edge cases—all of which may have contributed to improved LLM performance. For example, we include P12's full initial and final instructions in Appendix~\nameref{app:p12}. P12's final instructions contain multiple edge cases and provide a decision boundary, with more detail than their initial prompt.
This refinement suggests that \ourmethode{} may support deeper understanding of the target concept. 

\subsection{Study 2}
Here, we assume that we do have a prior specification of the target concept
and we would like to approximate the setting where a researcher's goal is to instruct an LLM 
to produce outputs which are grounded in that specification.
That is, the LLM should produce outputs which align with the ground truth labels. 
Thus, in addition to the steps conducted in \sone, 
in \stwo, we provide \subs{} with a training quiz which is designed to align their interpretation of \climate{} 
with the interpretation of the research team. 

The quiz consists of six videos. Initially, \subs{} are
shown \defnref{def1}. For each video they are asked 
to determine whether it contains \climate{} or not. 
They are then given feedback. If they are correct the feedback reinforces 
that decision and provides supporting arguments. 
If they are incorrect, they are provided with an explanation 
of why the research team assigned the label that they did. 
We see that this training quiz produced alignment with the research team; 
in \sone{} the accuracy of \subs{} on the test set was 53\% whereas it was 73\% in \stwo. 
Note, we kept the size of the test set small to limit the overall length of the study.
Given that \subs{} needed to perform many tasks within the study timeframe, we decided that labeling 10 videos was feasible. 

In \stwo, our research questions are around the ability of \ourmethode{}
to produce outputs which align with the team's labels. 
First, we investigate the impact of human involvement. 
That is,
 \textbf{RQ2:} How do the outputs produced by the LLM using instructions that \subs{} construct with \ourmethode{} compare with those produced using only the system prompt combined with \defnref{def1}?  
Similarly, we compare to an automated method which improves prompts without human intervention and ask, \textbf{RQ3:} How do the outputs produced by the LLM using instructions that \subs{} construct with \ourmethode{} compare with those produced using the prompts written by automated improvement methods?
Then we ask, \textbf{RQ4:} Can \subs{} use \ourmethode{} to improve the ability of LLMs to align with the team's labels?

Across these questions we use the same notation as in evaluating \sone. 
That is, we generate a set of labels from a LLM supplied with  instructions written  by participant $j$ in the $k$-th iteration,  $y^k_j$. 
To calculate the average change in accuracy between the final and initial instructions for \stwo, we compute the following: 

$$\frac{1}{|J|}\sum_{j\in J}(Accuracy(y_j^k,y_j^*)-Accuracy(y_j^0,y_j^*)),$$
where here we consider the ground truth $y_j^*$ to be the labels assigned by the research team. 

\begin{table*}[t]
\small
\caption{We compare the performance LLM outputs when produced with different instructions. 
For each set of outputs, we extract the labels and compare these to a set of ground truth  labels produced by the research team. 
We bold all differences which are statistically significant (p $<$ 0.05 according to Wilcoxon signed-rank tests).
In \stwo{} \subs{} selected their favorite prompt which we use below. We use the favorite prompt rather than the final prompt 
after receiving feedback in \sone{} that the best prompt was generally found several iterations before termination. 
}

\label{tab:prompt_performance24}
\begin{tabular*}{\textwidth}{@{\extracolsep{\fill}}ccccc}
\toprule
\multicolumn{5}{c}{RQ2: How do instructions \subs{} write with our \ourmethode{} compare with the system prompt?}\\
 \multicolumn{5}{c}{$\Delta=$ $\big($performance of favorite prompt$\big)$ $-$ $\big($performance of system prompt$\big)$}\\
\midrule
\textbf{$\Delta$Acc} & \textbf{$\Delta$Prec} & \textbf{$\Delta$Recall} & \textbf{$\Delta$F-Measure} & \textbf{$\Delta$ Avg Prompt Length}\\
 \textbf{0.531} & \textbf{ 0.172} & -0.061 & \textbf{ 0.217} & NA\\
\midrule
\multicolumn{5}{c}{RQ3: How do instructions \subs{} write with our \ourmethode{} compare with a state-of-the-art automated method?}\\
 \multicolumn{5}{c}{$\Delta=$ $\big($performance of favorite prompt$\big)$ $-$ $\big($performance of APE$\big)$}\\
\midrule
\textbf{$\Delta$Acc} & \textbf{$\Delta$Prec} & \textbf{$\Delta$Recall} & \textbf{$\Delta$F-Measure} & \textbf{$\Delta$ Avg Prompt Length}\\
 \textbf{0.234} & \textbf{0.123} & -0.011 & \textbf{0.147} & 453.0\\
\midrule
\multicolumn{5}{c}{RQ4: Can participants use
\ourmethode{} to improve their instructions to LLMs?}\\
\multicolumn{5}{c}{$\Delta=$ $\big($performance of favorite prompt$\big)$ $-$ $\big($performance of initial prompt$\big)$}\\
\midrule
   \textbf{$\Delta$Acc} & \textbf{$\Delta$Prec} & \textbf{$\Delta$Recall} & \textbf{$\Delta$F-Measure} & \textbf{$\Delta$ Avg Prompt Length}\\
  \textbf{0.185} & \textbf{0.090} & -0.010 & \textbf{0.108} & 461.6\\
\bottomrule
\end{tabular*}
\end{table*}

In \tabref{tab:prompt_performance24} we see positive results with respect to all three questions, when considering accuracy, precision, and F-Measure. 
With respect to \textbf{RQ2}, we see in \tabref{tab:prompt_performance24} that there is substantial improvement when using \ourmethode{} compared to the system prompt coupled with \defnref{def1}. That is, there is a clear need for human intervention in writing instructions for this task. 
While recall decreases, this is due to the fact that the system prompt outputs \climate{} much more often than the research team (100\% of the time vs 10\% of the time) for the 10 videos.

To evaluate \textbf{RQ3},
we compare against \emph{Automatic Prompt Engineer (APE)} \citep{ape_zhou2023}. APE treats natural-language instruction as a program and performs black-box optimization: an LLM proposes candidate instructions from input–output demonstrations; the target model scores candidates on a chosen metric; and the system iteratively filters and paraphrases high-scoring prompts before returning the top instruction. We see in \tabref{tab:prompt_performance24} that \ourmethode{} improves over APE in terms of accuracy, precision, and F-Measure. 
The recall is slightly lower due to the fact that the prompts produced by APE assign 72.05\% positive labels.

To answer \textbf{RQ4} we compare the performance between the favorite prompt and the initial prompt. 
In \tabref{tab:prompt_performance24}, the performance of the prompts provided with instructions written after 
\subs{} used \ourmethode{} is higher than that of the initial prompts. The improvement of the accuracy, precision, and F-Measure are all statistically significant. The length of \subs' instructions also increased substantially over time, with initial instructions averaging 381.7 characters and final instructions averaging 843.3.


\begin{figure}[h]
  \centering
  \includegraphics[width=.8\linewidth,clip]{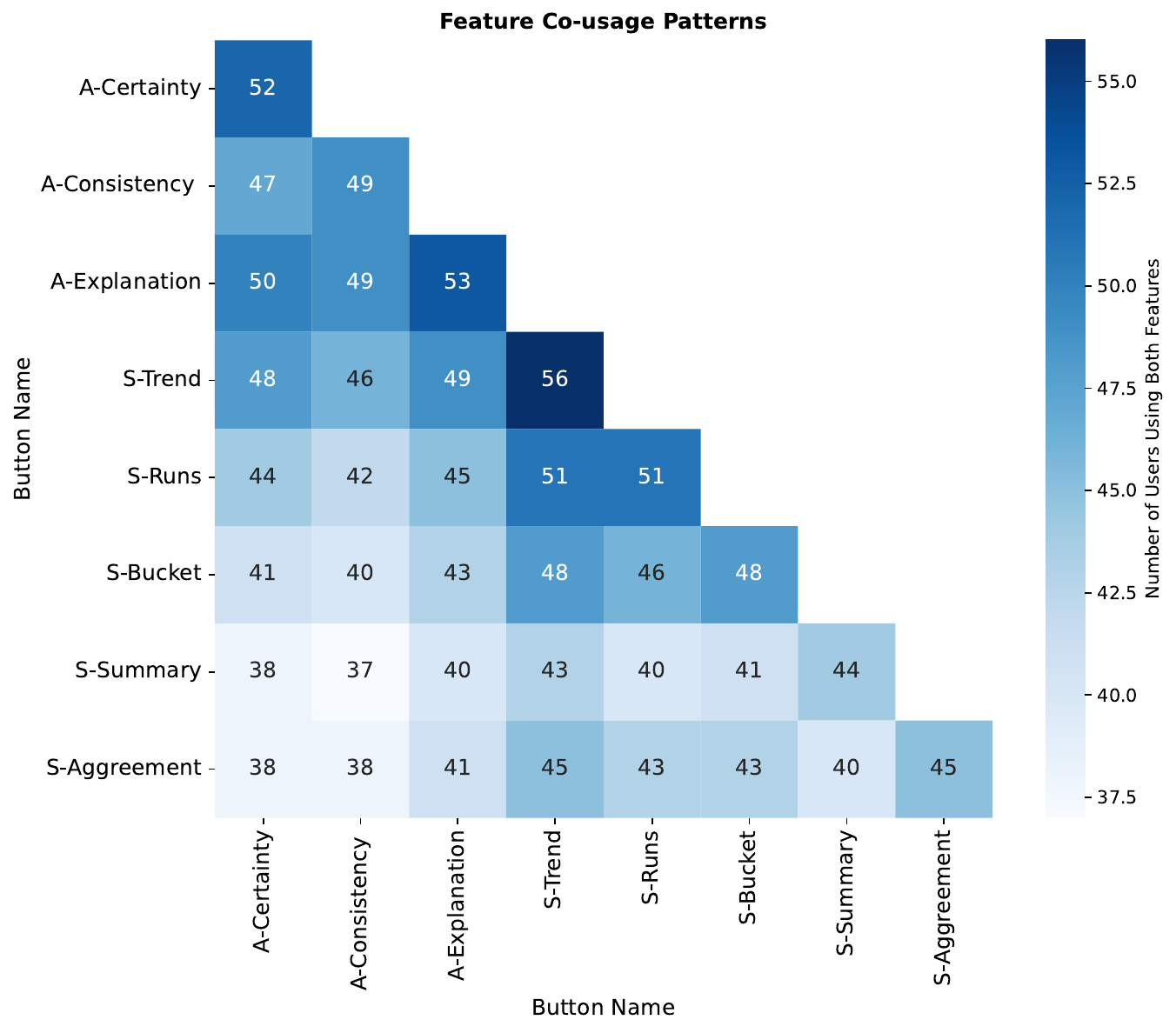}
  \caption{The number of times features were used. By reading the diagonal we can see that most of the 60 \subs{} used all features, with S-Trend being used the most often.}
  \label{fig:feature_co_usage_heatmap}
\end{figure}

\subsection{\ourmethode{} usability}
Additionally, we are interested in how \subs{} used \ourmethode, 
and if this generally non-technical population found the tool to be
usable. In \figref{fig:feature_co_usage_heatmap} we see which features were selected with which other features. 
In particular, \subs{} tended to use the \annexpl, \anncert, and \statren, features the most. 
In \sone{} we also asked \subs{} to respond to a free-text question asking which features they found most helpful. 
There, we found that \subs{} most frequently mentioned \annexpl. 
In \stwo{} we updated this question so that \subs{} where asked to rate each of the 
information features on a scale from ``not at all helpful'' to ``extremely helpful''. 
We find that \subs{} rated \annexpl{} as the most helpful on average, scoring 4.36 on a 1 to 5 scale (see \figref{fig:feature_helpfulness_scatter}). 





\begin{figure*}
  \centering
  \includegraphics[width=.59\linewidth]{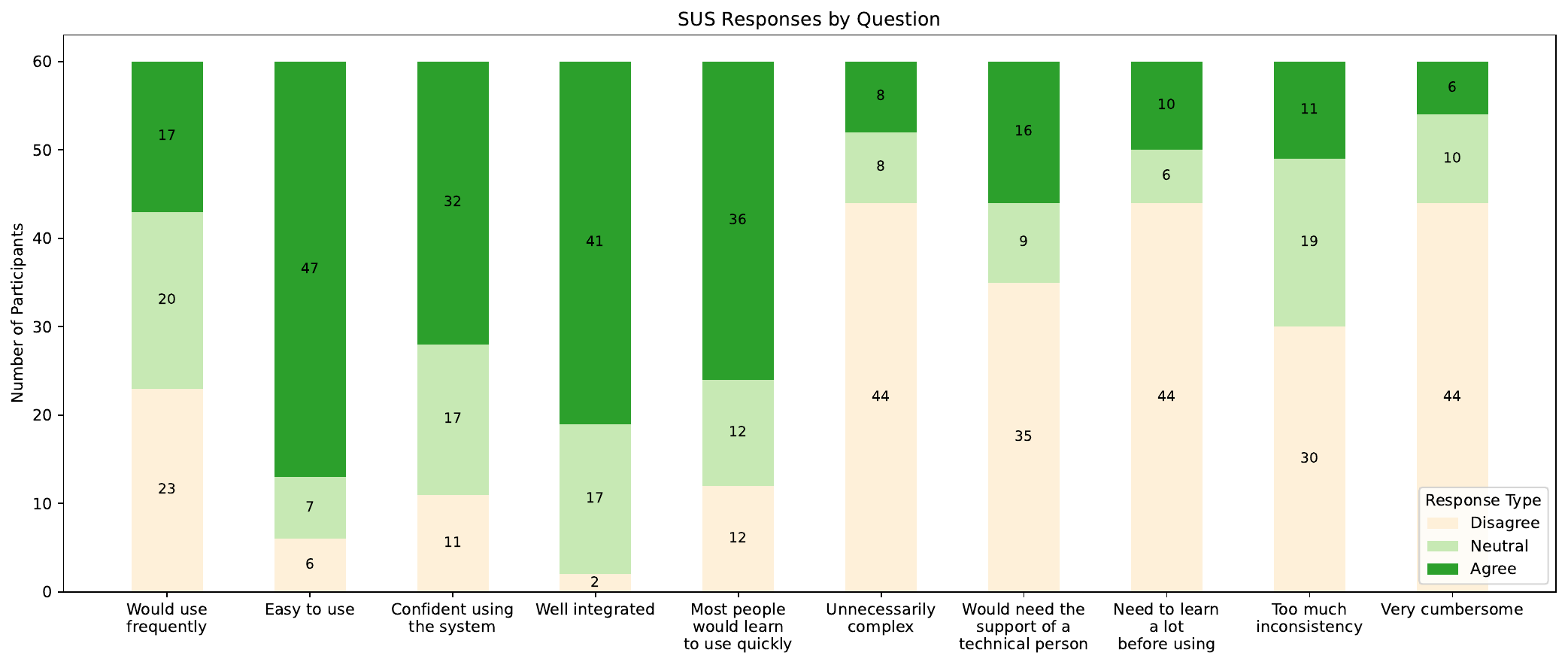}
  \caption{Participants completed the SUS  with respect to \ourmethode.
    We see that overall the system is usable and does not require a technical background. The numbers in the bars refer to the number of participants who selected this option.}
  \label{fig:sus_stacked_barplot}
\end{figure*}

Overall, participants responded positively to \ourmethode. The average System Usability Scale (SUS) score across all participants was 67, and the median was 68, with the 75 percentile scoring at 75 indicating high perceived usability. According to standard SUS interpretation, scores above 68 are considered above average, and scores above 80 are typically considered excellent.
Specific ratings for each SUS question can be found in \figref{fig:sus_stacked_barplot}. 



\section{Discussion}
We find that \ourmethode{} supports 
\subs' interactions with LLMs in two contexts. 
Whether they have an existing \spec{}  or not, 
\subs{} can use \ourmethode{} to improve \align{} in that concept. 


The method we propose is only one component of a full annotation pipeline. 
We limited the manual annotation task to 10 videos because our six-step workflow requires time for participants to iteratively refine instructions with the system in addition to labeling within a single ($\leq$ 2-hour) session, this tradeoff constrained the number of videos we could feasibly annotate.
Other work could apply \ourmethode{} in a broader annotation pipeline, for example to improve supervised fine-tuning \citep{pangakis2025keeping}. Here, 
a researcher would be in the second setting where the concept specification exists, 
and their primary goal is to design the best prompt for their setting. 
Our method can also be useful for non-technical scientists to learn more about 
LLMs through an example setting in which they have rich expertise. 
Through engagement with the information features, 
they can analyze LLM reliability, fidelity, and quality.

Our work also speaks to the literature on overreliance. 
We saw in \sone, that \subs{} tended to change their labels before and after using \ourmethode{}
in the direction of the LLM outputs (see \figref{fig:majority_yes_count} in Appendix~\nameref{app:labelshift}).
We did not see this effect in \stwo{} where \subs{} were given additional training and grounding in the target concept 
and were shown expert labels produced by the research team in addition to LLM labels.
While preliminary, this suggests that  additional training and external anchors 
can act as counter-weights to LLM outputs.

Given that a potential use-case of our system is a researcher asking research assistants
to use it to design prompts (either for zero-shot, few-shot, or supervised learning), 
we sound a cautionary note that researchers should monitor the ways in which research 
assistants engage with LLMs in annotation pipelines. 
Providing expert / ground-truth anchor points on even a small number of examples 
can likely reduce the risks of overreliance.

\section{Related Work}
\label{sec:relatedwork}
Our work contributes to understanding around human-LLM interactions in annotation workflows. 
To ground our system in a real CSS use case, we consider the under-studied problem of \commentout{\gs{eek.  Maybe do global find and replace.  Need "mitigation".} \zx{updated}}climate change mitigation pessimism on social media. 
We then discuss existing systems for using LLMs for annotation, including prompt engineering tools which are related to our approach for instruction design. 
Next, we discuss challenges with LLM annotations. Finally, we summarize our contributions to related work.

\subsection{Social media and responses to climate change}
Many researchers have studied attitudes and beliefs around climate change on social media platforms, with a bias towards the platform which was previously Twitter \citep{pearce2019social}. 
Existing work has studied broad discussion areas around climate change on social media platforms \citep{dahal2019topic,bohm2025exploring,qiao2022topic,chen2023climate}. 
To the best of our knowledge no one has investigated attitudes around \textit{mitigation} in a data-driven way before. 

\commentout{Often, understanding more nuanced and specific questions requires hand-coding and close inspection of a manageable amount of documents \citep{bloomfield2019circulation,walter2018echo,jacques2016hurricanes,cann2018does}. To understand how non-governmental organizations frame climate change on Facebook, a team hand-coded a few hundred posts \citep{vu2021social}. Following the approach we describe in the introduction, a faculty member supervised graduate students who performed the coding. 
While these studies offer more detailed analysis than the automatic topic modeling approach, they do not scale \citep{metag2016content}. 
Our approach aims to utilize the oversight of trained RAs to detect specific nuanced concepts and offer scalability.}


\subsection{Prompt engineering tools}
To support non-technical users, such as social scientists, a growing body of work has explored systems for LLM-based annotation that enable prompt engineering. Tools like APT-Pipe aim to optimize prompt quality with minimal human labor by selecting examples for few-shot prompts after the initial design \citep{zhu_apt-pipe_2024}. Others, such as ChainForge and visual prompt editors, provide interfaces for testing prompt variations and visualizing LLM behavior \citep{arawjo2024chainforge, strobelt_interactive_2023}. Recent work further shows prompt design can substantially affect LLM annotation quality \citep{atreja2025}. Automatic Prompt Engineer (APE) goes further by automating instruction search via an LLM-driven propose–score–select loop over few-shot demonstrations \cite{ape_zhou2023}. We adopt APE as a strong automated state-of-the-art comparator with \ourmethode{}. 
However, these systems often rely on users supplying ``gold-standard'' labels or are primarily designed for NLP experts. In contrast, many real-world experts, such as those in the social sciences, are more likely to require support in interpreting ambiguous or contested concepts rather than writing formal prompt templates.
Non-AI experts often struggle to craft robust prompts, explain prompt effects, and to evaluate them systematically: a study with a no-code prompt-design tool found that participants explored prompts opportunistically, over-generalized from single successes or failures, and rarely conducted systematic testing \citep{zamfirescu2023johnny}. 
Our work is firmly human-centered and we show  that our approach improves over automated methods for a particular difficult CSS setting. In future work we will expand to additional settings. 



\subsection{Challenges with LLM annotations}  
While LLMs are increasingly used as scalable alternatives to human labeling \citep{gilardi_chatgpt_2023}, they introduce significant challenges when applied to subjective social science tasks.
Recent evaluations show that zero‑shot LLM annotations still fall short on nuanced tasks and can mislead downstream analyses \citep{lim2024rigor}, and automated LLM annotation performance is subpar on a large number of tasks \citep{pangakis2025keeping}.
In particular, the poor performance of zero-shot LLMs on a number of tasks demonstrates the need for keeping humans in the annotation loop \citep{pangakis2025keeping}.
Beyond general performance, relying on LLMs also risks codifying systematic errors, recent studies demonstrate that classifiers trained on LLM data can propagate severe racial biases \citep{okpala2025large}, and persona-based LLMs often fail to align with the actual socio-demographic nuances of human annotators\citep{giorgi2025human}. We draw on these strategies by providing both explanations and uncertainty metrics, 
two techniques which have been found to be successful in the past \citep{raees2025exploring}.
Additionally, we introduce a new information feature, that of the social science expert annotation, 
which serves as a counter-point to the AI annotation. 

Together, our information features
provide \subs{} with opportunities to think analytically, and to critique the AI 
when it proves to be unreliable or in contradiction to human expertise. 
Like existing work, we utilize natural language explanations \citep{wang2024human}. 
However, there is a potential that such explanations may lead \subs{} to falsely believe that LLMs 
are interpretable while they are not \citep{rudin2021interpretable} (that is they are not ``glass boxes''). 
Rethinking interpretability around LLMs 
is an active area of research \citep{singh2024rethinking}, 
and we designed \ourmethode{} to be a useful foundation for such research.

\commentout{These validity and bias issues exacerbate the risks associated with human-AI collaboration. A known problem with human-AI collaboration is \textit{overreliance},
when human users (such as annotators) agree with LLM outputs, 
even when the LLM is incorrect \citep{buccinca2021trust}. 
Overreliance can lead to human users making worse decisions with AI
than they would on their own, a truly undesirable outcome in human-AI systems
 which undermines  complementary team performance \citep{bansalctp2021}. 
For example, there is evidence of overreliance in clinical  \citep{jacobs2021machine},
academic \citep{lai2020chi}, and financial settings \citep{klingbeil2024trust}, 
among others \citep{zhang2020effect,prabhudesai2023understanding}. 

Effective strategies for reducing this phenomenon include:
those which nudge a person to perform analytic thinking (cognitive forcing functions)
\citep{buccinca2021trust};
supplying explanations of an AI's decision in a way which reduces the difficulty of achieving the task relative to the absence of AI assistance
\citep{vasconcelos2023cscw}; 
and personalizing explanations to particular human users \citep{raees2025towards}. 
}

\subsection{Contributions to Related Work}
In contrast to fully-automated LLM annotation pipelines, we propose an approach which 
builds on the rich tradition of iterative mentorship in the social sciences. 
Our system provides \subs{} with many opportunities to reflect analytically on LLM behavior, 
in order to improve their instructions over time. 
We test our system on the societally-relevant task of \climate{} detection, 
thus contributing to the field of understanding public responses to climate change
 through social media data. 
Our approach blends the strength of two common analysis strategies which either utilize focused human effort to annotate small collections of documents, 
 or computational approaches which tend towards broader categorizations of sentiment;
 we use computational methods and human effort to detect a specific concept with scientific relevance. 
Looking forward, we envision systems like \ourmethode{} supporting annotation workflows in CSS settings, either when a concept is specified or when it is not.



\bibliography{AnonymousSubmission/LaTeX/aaai2026}



\appendix

\section{Data Storage}
Participants will upload data to the institution Google Drive. From there, we will backup the data to an institution data center. We do not share any of our data.

The data center is protected through multiple layers of physical security and layered network protection firewalls. The hardware is monitored by a professional technical team running continuous monitoring software to detect malicious access, running automatically scheduled updates, and cooperating with the campus computing teams to keep abreast of potential and active security threats.

\section{~\tabref{tab:sone_performance} Full Data}
\label{app:tab2full}

The full data used to calculate~\tabref{tab:sone_performance} is shown in~\tabref{tab:sone_performance_full}.

\begin{table*}[t]
\small
\caption{For \sone{}, we inspect how well the LLMs perform on their first and favorite prompts against \subs{} labels before and after using \ourmethode{}. 
Metrics: accuracy, precision, recall, F1. The average and standard error of the mean (shown next to respective metrics in footnote size) are computed across three random seeds over three runs with gpt-4o-mini-2024-07-18, at temperature 0.7.
} 
\label{tab:sone_performance_full}
\begin{tabular*}{\textwidth}{@{\extracolsep{\fill}}lcccccc}
\toprule
Ground Truth from & Labels Produced by & \textbf{Acc} & \textbf{Prec} & \textbf{Recall} & \textbf{F-Measure} & \textbf{Average  Length of Instructions}\\
\midrule
step 3 & first prompt & $0.594_{0.010}$ & $0.547_{0.008}$ & $0.810_{0.010}$ & $0.626_{0.008}$ & 236.7\\
step 3 & last prompt & $0.657_{0.007}$ & $0.629_{0.007}$ & $0.715_{0.024}$ & $0.645_{0.012}$ & 632.3\\

\midrule
step 5 & first prompt & $0.563_{0.008}$ & $0.561_{0.005}$ & $0.806_{0.004}$ & $0.619_{0.006}$ & 236.7\\
step 5 & last prompt & $0.678_{0.002}$ & $0.673_{0.005}$ & $0.734_{0.019}$ & $0.668_{0.010}$ & 632.3\\

\midrule
\end{tabular*}
\end{table*}

\section{Prompt Performance on Participants' Labels for \stwo{}}
\label{app:subss2}

A replication of~\tabref{tab:sone_performance_full} is shown in~\tabref{tab:two_performance_full} to inspect how the LLMs perform on \subs{}'s first and favorite prompts against participants labels before and after using \ourmethode{} for \stwo{}.

\begin{table*}[t]
\small
\caption{We repeat~\tabref{tab:sone_performance_full} for \stwo{}. We inspect how well the LLMs perform on their first and favorite prompts against \subs{} labels before and after using \ourmethode{}. 
Metrics: accuracy, precision, recall, F1. The average and standard error of the mean (shown next to respective metrics in footnote size) are computed across three random seeds over three runs with gpt-4o-mini-2024-07-18, at temperature 0.7.
} 
\label{tab:two_performance_full}
\begin{tabular*}{\textwidth}{@{\extracolsep{\fill}}lcccccc}
\toprule
Ground Truth from & Labels Produced by & \textbf{Acc} & \textbf{Prec} & \textbf{Recall} & \textbf{F-Measure} & \textbf{Average  Length of Instructions}\\
\midrule
step 3 & first prompt & $0.522_{0.009}$ & $0.356_{0.006}$ & $0.650_{0.012}$ & $0.425_{0.007}$ & 381.7\\
step 3 & favorite prompt & $0.628_{0.008}$ & $0.402_{0.016}$ & $0.556_{0.015}$ & $0.420_{0.011}$ & 843.3\\

\midrule
step 5 & first prompt & $0.518_{0.009}$ & $0.328_{0.004}$ & $0.636_{0.005}$ & $0.389_{0.004}$ & 381.7\\
step 5 & favorite prompt & $0.636_{0.004}$ & $0.392_{0.007}$ & $0.532_{0.018}$ & $0.399_{0.012}$ & 843.3\\

\midrule
\end{tabular*}
\end{table*}

\section{~\tabref{tab:prompt_performance24} Full Data}

The full data used to calculate~\tabref{tab:prompt_performance24} is shown in~\tabref{tab:raw_bigtable}.

\begin{table*}[t]
\small
\caption{
We show the average performance across participants using their first vs. favorite prompt in \stwo{}. 
Metrics: accuracy, precision, recall, F1. The average and standard error of the mean (shown next to respective metrics in footnote size) are computed across three random seeds over three runs with gpt-4o-mini-2024-07-18, at temperature 0.7.
For each set of outputs, we extract the labels and compare these to a set of ground truth labels produced by the research team.
We also show the performance of a state-of-the-art method APE. 
In the system prompt scenario, the following basic prompt that is the same definition given to participants for manual annotation replaces participant prompt: ``Do you think the following video displays pessimism with regard to climate change mitigation (the ability to take action to reduce the event of and effects of climate change)?''
} 
\label{tab:raw_bigtable}
\begin{tabular*}{\textwidth}{@{\extracolsep{\fill}}lcccccc}
\toprule
 Labels Produced by & \textbf{Acc} & \textbf{Prec} & \textbf{Recall} & \textbf{F-Measure} & \textbf{Average Prompt Length}\\
\midrule

\multicolumn{6}{c}{System Prompt}\\
\midrule
system prompt & $0.100_{0.000}$ & $0.100_{0.000}$ & $1.000_{0.000}$ & $0.182_{0.000}$ & -\\

\midrule
\multicolumn{6}{c}{\stwo{} - \ourmethode{}}\\
\midrule


first prompt & $0.446_{0.008}$ & $0.182_{0.003}$ & $0.949_{0.008}$ & $0.292_{0.001}$ & 381.7\\
favorite prompt  & $0.631_{0.006}$ & $0.272_{0.012}$ & $0.939_{0.000}$ & $0.400_{0.010}$ & 843.3\\
\midrule


\multicolumn{6}{c}{SOTA Comparator}\\
\midrule
APE & $0.397_{0.002}$ & $0.149_{0.000}$ & $0.950_{0.008}$ & $0.253_{0.001}$ & -\\
\bottomrule
\end{tabular*}
\end{table*}

\section{System Prompt}
\label{app:systemPrompt}
\footnotesize

\texttt{Provide your 2 best guesses and the probability that each is correct (0.0 to 1.0) for the following question. }

\texttt{Give ONLY the guesses and probabilities, and include an explanation for the most likely guess only after specifying "Explanation:" in the end. For example:} 

\texttt{G1: first most likely guess, Yes or No answer to the question! }

\texttt{P1: the probability between 0.0 and 1.0 that G1 is correct, without any extra commentary whatsoever; just the probability! ... }

\texttt{G2: 2-th most likely guess, Yes or No answer to the question! }

\texttt{P2: the probability between 0.0 and 1.0 that G2 is correct, without any extra commentary whatsoever; just the probability! }

\texttt{Explanation: include your explanation here. }

\texttt{The question is: [Participant Instructions will be appended here]}


\section{Label Shift} 
\label{app:labelshift}

\begin{figure}
  \centering
  \begin{subfigure}[t]{0.48\textwidth}
  \includegraphics[width=\textwidth]{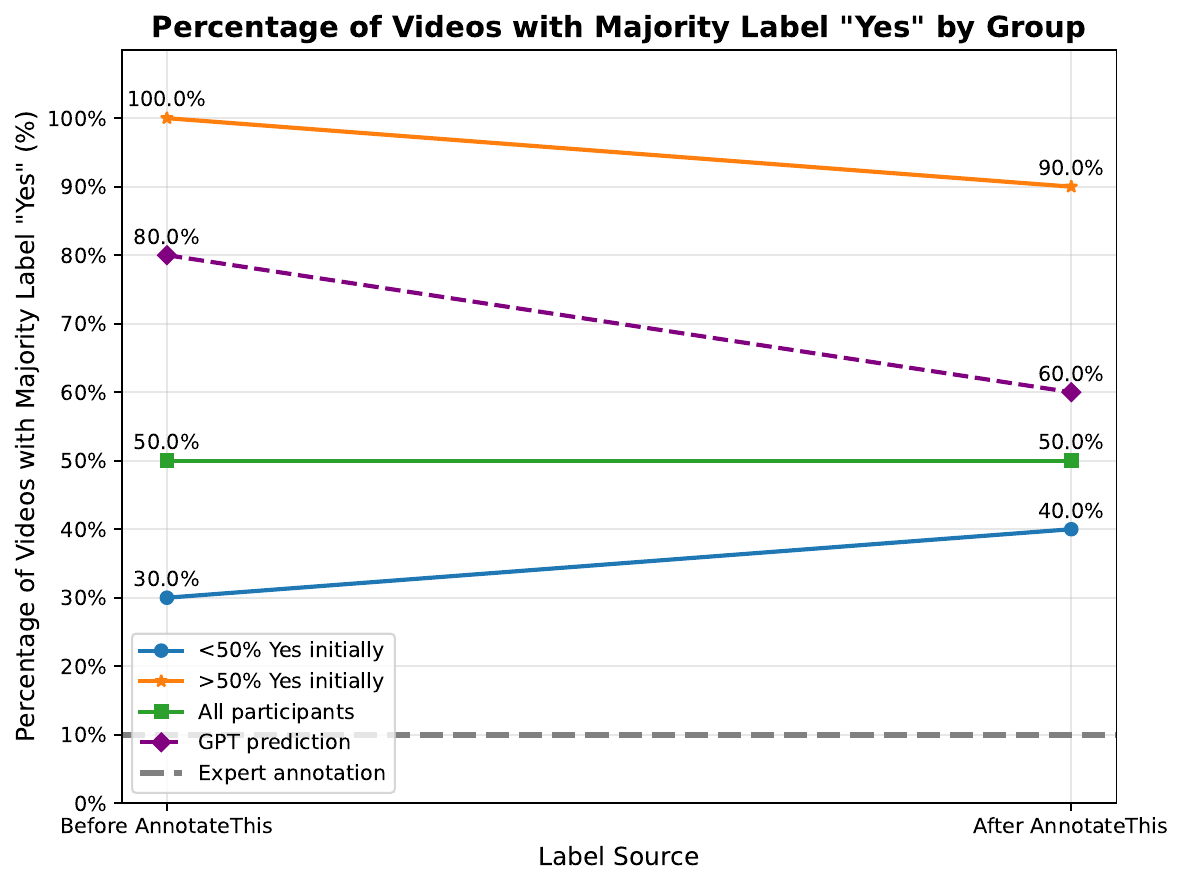}
  \caption{\sone: We see that \subs{} who begin by assigning more than 50\% of videos the label ``Yes''
  assign fewer ``Yes'' labels after using \ourmethode{} (100\% to 90\%). This is generally a positive sign, as they are overall assigning far more positive labels than the social science experts who only assign the positive label to 10\% of videos. However, we also see that \subs{}
  who begin by assigning fewer than 50\% of videos the label ``Yes'' actually assign the label more often after using \ourmethode{} (30\% to 40\%). This suggests that they may be influenced by the LLM which tends to assign the positive label often (initially 80\% and then 60\% of the time). 
  }
  \label{fig:majority_yes_count}
  \end{subfigure}
    \hfill
  \centering
  \begin{subfigure}[t]{0.48\textwidth}
  \includegraphics[width=\textwidth]{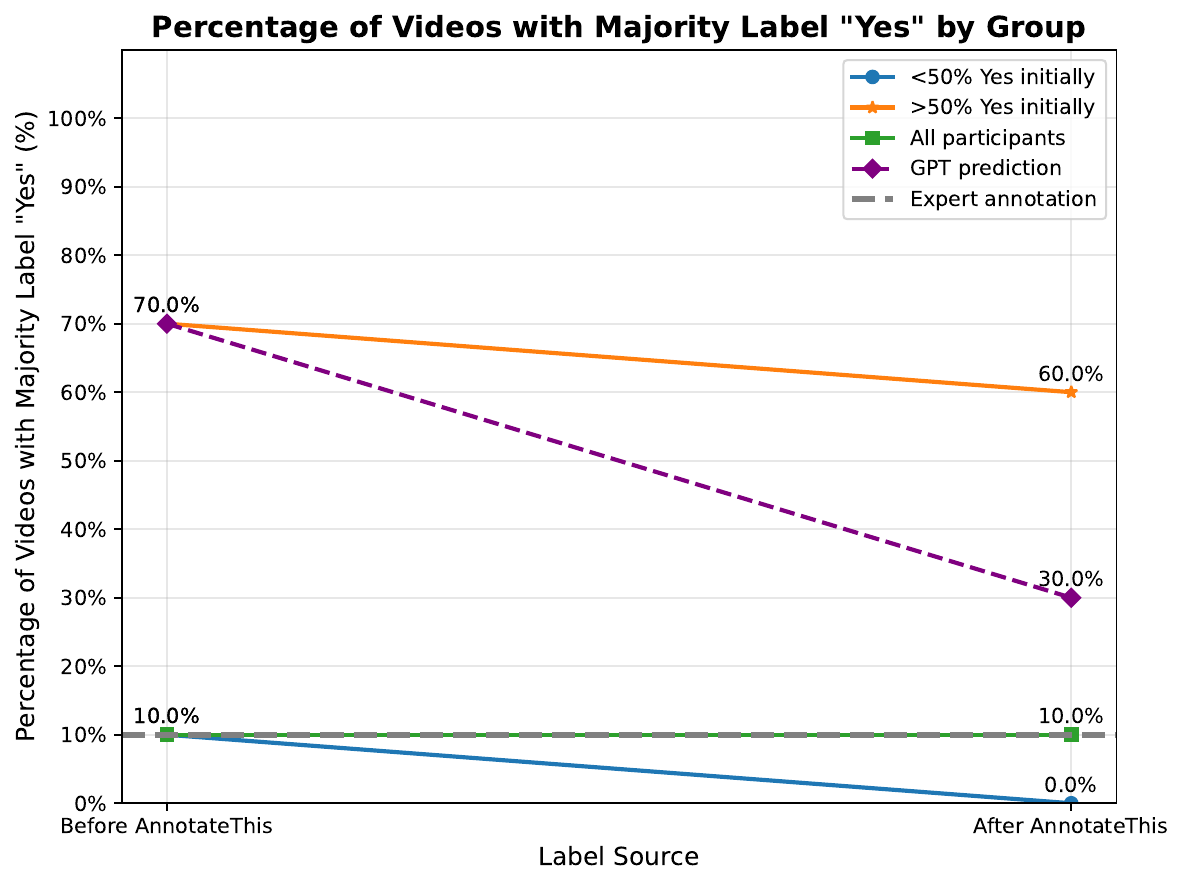}
  \caption{\stwo: We see that \subs{} who begin by assigning more than 50\% of videos the label ``Yes''
  assign fewer ``Yes'' labels after using \ourmethode{} (70\% to 60\%). This is generally a positive sign, as they are overall assigning far more positive labels than experts who only assign the positive label to 10\% of videos. In \stwo{} we see that \subs{}
  who begin by assigning fewer than 50\% of videos the label ``Yes'' assign the label \textbf{less} often after using \ourmethode{} (10\% to 0\%). In contrast to \sone, here, \subs{} do not seem to be influenced by the LLM which tends to assign the positive label often (70\% to 30\%).  There are several differences between \stwo{} and \sone. The LLM's behavior is directed towards experts more effectively here. Also, on average, \subs{} assign the label far less often than in \sone, in a much clearer alignment with the frequency with which it is assigned by experts.
   }
  \label{fig:majority_yes_count2}
  \end{subfigure}
\caption{Percentage of ``Yes'' labels for different groups. 
  Participants are asked to annotate the videos themselves twice: once before using \ourmethode{} (Before \ourmethode{} in the figure) and once after (After \ourmethode{} in the figure). 
  \label{fig:both}
  }
\end{figure}

The figures showing participants' label percentage shifts from step 3 and step 5 are shown in \ref{fig:both}.

\section{Feature Helpfulness} 
\label{app:fhelpfulness}

\begin{figure}[t] 
  \centering
  \includegraphics[width=.9\linewidth]{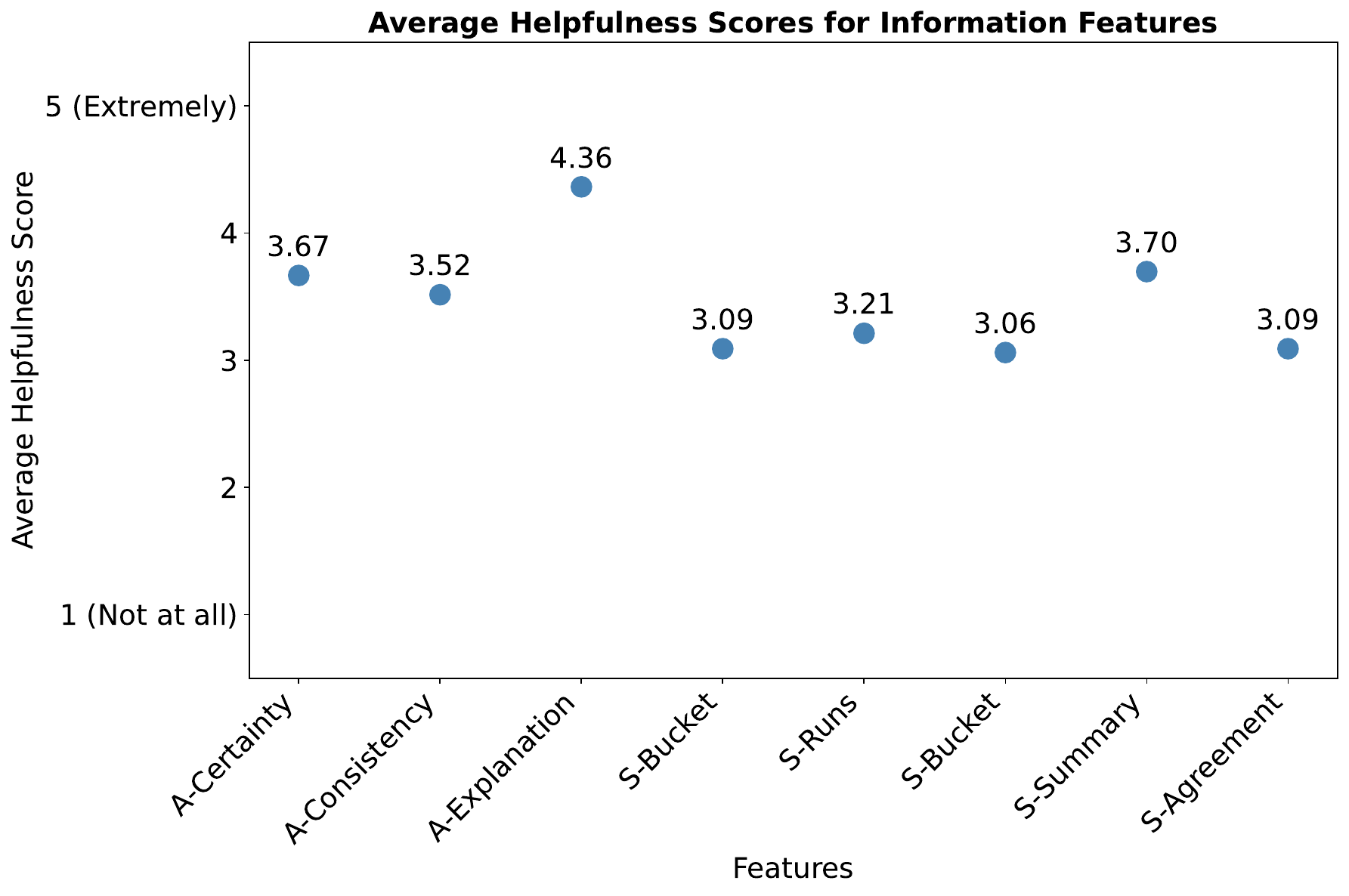}
  \caption{Participants in \stwo{} were asked to rate the helpfulness of each feature. We see that the features that included natural language tended to be viewed as more helpful than the statistical features.}
  \label{fig:feature_helpfulness_scatter}
\end{figure}

The figure showing participants' rating for feature helpfulness in \stwo{} is shown in Fig~\ref{fig:feature_helpfulness_scatter}.

\section{Participant Background}
\label{app:participants}
In this section, we provide more context on the 60 participants recruited (27 for \sone{} and 33 for \stwo{}). The participants were recruited through departmental mailing lists and digital posters (displayed physically on monitors in public spaces in a departmental building). 

\paragraph{\textbf{Background and technical fluency}}
39 (\(65.0\%\)) of the participants reported that they had prior experience coding data or analyzing media content: open-ended survey responses revealed prior use of R or Python for coursework (e.g.\ statistical coding, map making, API retrieval) and qualitative coding software such as NVivo. Only a small number of 8 (13.3\%) participants had previously trained human annotators. This indicates that a considerable portion of our participants have certain experience with coding data or annotation pipelines, and would potentially benefit from a tool like \ourmethode{} in their future work.

\paragraph{\textbf{Familiarity with LLMs}}
35 participants (\(58.3\%\)) reported frequent use of ChatGPT or similar tools, while the remaining 22 (\(36.7\%\)) had merely ``experimented’’ with such models but did not use them regularly, and only 3 (\(5.0\%\)) of the participants selected that they have never used ChatGPT or other tools. The majority of the participants (\(75.0\%\)) have no prior experience training large language models to analyze text data, and only 7 of them (\(11.7\%\)) indicated that they had such experience, confirming that \ourmethode{} represented a relatively novel workflow for the cohort.

\paragraph{\textbf{Domain Expertise and Climate Attitudes}}
Our participants primarily consisted of master's students (91.7\%), with a smaller representation of 2 (3.3\%) PhD students and 3 (5.0\%) faculty or staff. Participants also expressed substantial concern about climate change: the majority were moderately to highly worried (83.3\%) selecting 4-5 on a scale of 5. However, views on climate change mitigation pessimism — the belief that ``the world will be unable to mitigate the most serious consequences of climate change'' — were more varied: while 35.0\% leaned toward agreement (selecting 1–2), 36.7\% selected 4, and 18.3\% selected 5, reflecting a range of outlooks within the group. This distribution underscores that climate change mitigation pessimism is both a contested and deeply relevant concept, making it a timely and meaningful focus for interpretive annotation.

\paragraph{\textbf{Additional \stwo{} Background Question Responses}}
We included additional questions on participant engagement with social media (TikTok), and their educational background with LLMs in study 2. In their responses, 20 (60.6\% out of 33) participants reported that they have never used TikTok, 2 (6.1\%) use it a few times a month, and 6 (18.2\%) use it a few times a week, and 5 (15.2\%) use TikTok a few times a day. We then asked participants about their educational training on LLMs. 24 (72.7\%) of the participants have not taken any classes on this subject or studied this subject on their own, 7 (21.2\%) have taken one class on this subject and/or studied a bit on their own, and only 2 (6.1\%) have taken several classes and/or studied on their own. This is align with our goal of designing \ourmethode{} for a non-technical audience.

\section{Tasks}
\label{app:tasks}
Participant underwent a structured five-step workflow, where in \stwo, they completed one additional task:
\begin{itemize}
    \item \textbf{Step 1} Sign a consent form containing IRB approval information for the study with accompanying verbal and written instructions introducing the study and the task.
    \item \textbf{Step 2 (\stwo{} participants only)} Complete a training quiz. Participants first view a TikTok video, assign a label for the presence of \climate{} based on their own judgment, then view the expert label and explanation for the video, and repeat this process for 6 videos in total. 
    \item \textbf{Step 3} Manually annotate 10 TikTok videos for the presence of \climate{} based on \defnref{def1}, and their own judgment. They are provided a web form to view TikTok videos and enter their labels. We also supplied headsets for listening to the audio, and  gave instructions that participants finish the annotation at their own pace. Participants' initial labels served as baseline to infer their operationalization of the subjective target concept.
    \item \textbf{Step 4} Iteratively engage with \ourmethode{}. Participants are first directed to the \prm{} page (personal LLM-Instruction hub), where they write initial instructions. These instructions are used to generate LLM predictions on a fixed set of videos. After viewing the model’s predictions, participants can access the set of eight information features to help them interpret the results. Based on this feedback, participants can revise their instructions and rerun the model as many times as they like. The system tracks the number of clicks on each information feature to capture interaction patterns. At any point, participants may choose to proceed to the next task phase.
    \item \textbf{Step 5} Re-annotate the same 10 videos as in Step 2. This data is collected to compare with the annotations in Step 2 to investigate if and how RAs's annotations change.
    \item \textbf{Step 6} Complete a post-task survey. Immediately after participants finished their final instruction–iteration cycle, we administered a brief questionnaire combining SUS and custom questions. It captured perceived usability of \ourmethode{}, prior annotation and LLM experience, reasons for ending further iterations, and baseline attitudes toward \climate—context we use to interpret behavioral logs. 
\end{itemize}

\section{Keywords Selection Process.}
\label{app:keywords}

\begin{figure*}[t]
  \centering
  \includegraphics[width=\textwidth]{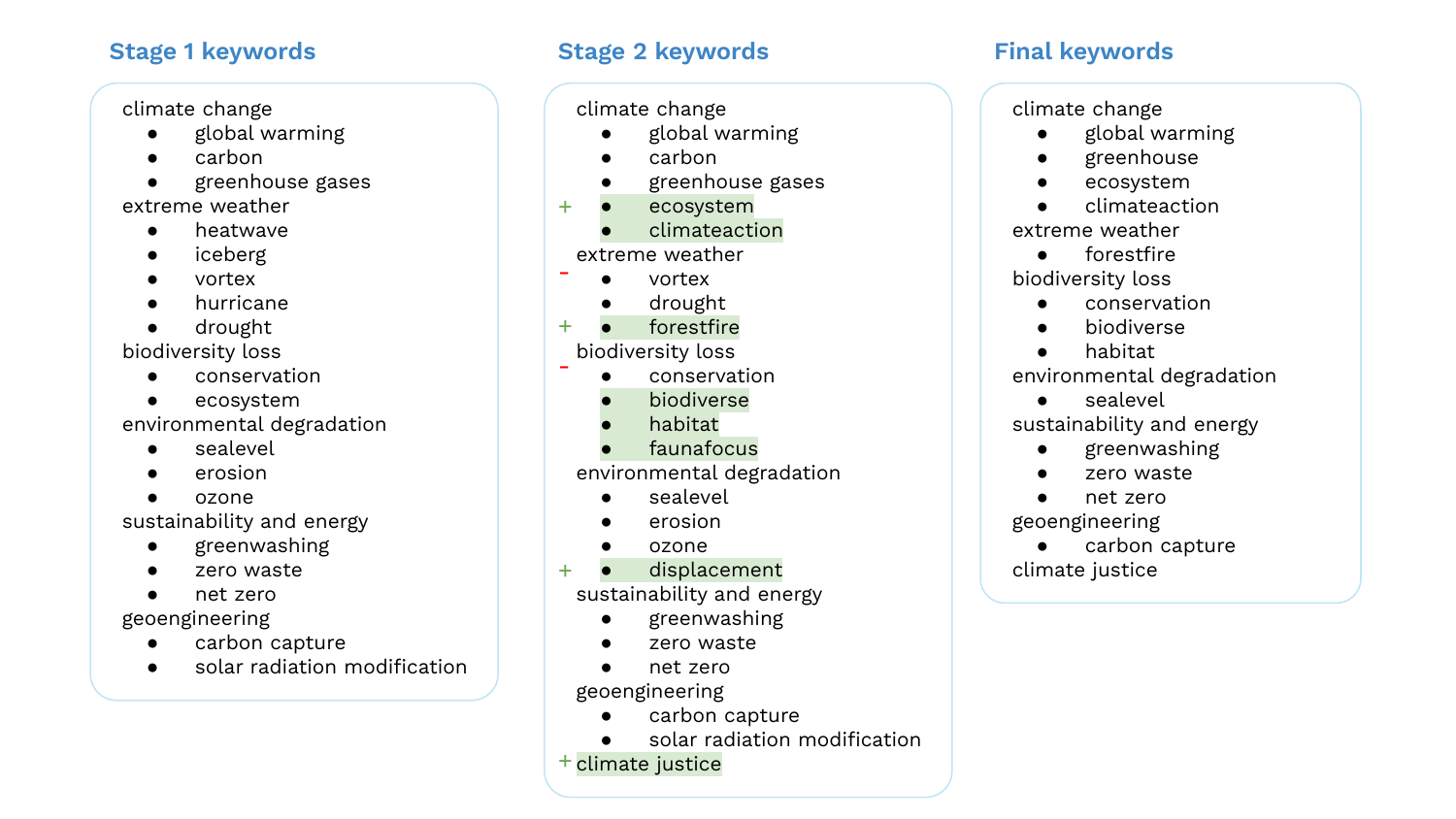}
  \caption{Keywords selection process. The team finalized the keyword set over two stages. Newly added keywords are highlighted; keywords removed from the previous stage are marked with a “$-$”.}
  \label{fig:keyword_selection}
\end{figure*}

We list all the keywords and their revisions during our two-stage keywords selection process in Fig~\ref{fig:keyword_selection}.

\begin{figure}[t]
  \centering
  \includegraphics[width=\textwidth]
  {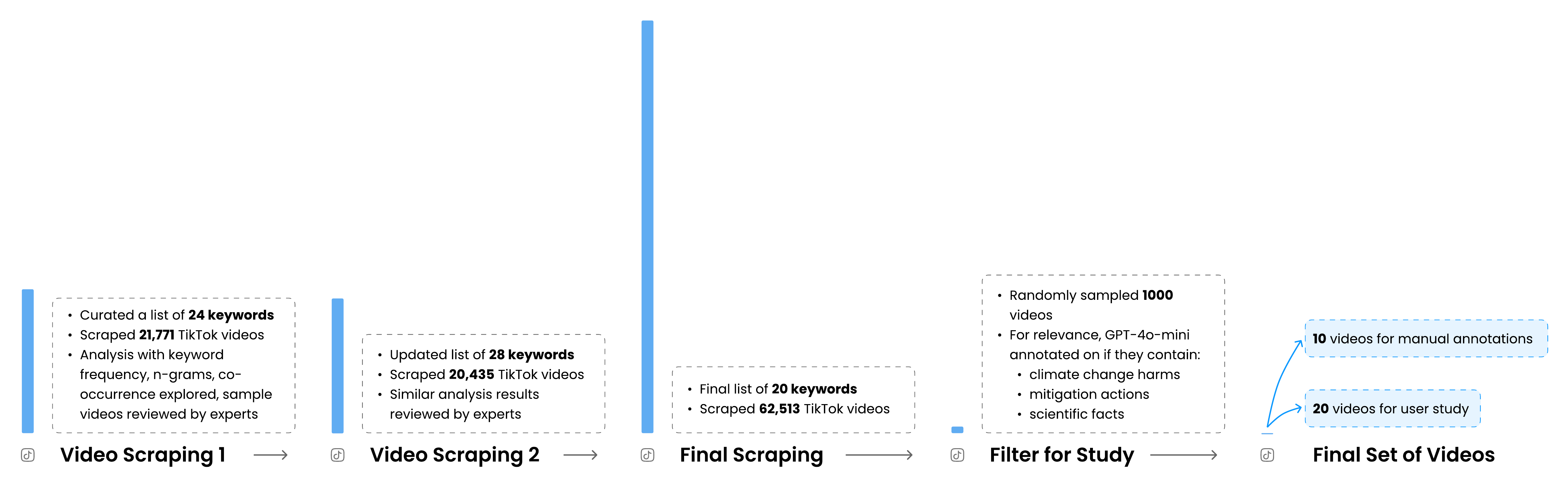}
  \caption{TikTok videos dataset curation process.}
  \label{fig:scraping}
\end{figure}

We describe our dataset curation process in more detail in Fig~\ref{fig:scraping}. 
The figure shows the complete five steps in the dataset curation process:
(1) \textit{Video Scraping Stage 1}: curated 24 keywords; scraped 21{,}771 TikTok videos; ran keyword, $n$-gram, and co-occurrence summaries; experts spot-checked samples. 
(2) \textit{Video Scraping Stage 2}: updated to 28 keywords; scraped 20{,}435 videos; repeated analyses and expert review. 
(3) \textit{Final Scraping}: finalized 20 keywords; scraped 62{,}513 videos. 
(4) \textit{Filter for Study}: randomly sampled 1{,}000 videos; GPT-4o-mini screened for relevance (climate-change harms, mitigation actions, scientific facts). 
(5) \textit{Final Set of Videos}: selected 10 videos for manual annotations, 50 videos sent to the LLM in \sone{}  and 20 videos sent to the LLM in \stwo.





\section{Example Instructions} 
The first and final instructions for participant P12 in \sone{} is shown in ~\tabref{tab:p12_prompts}.

\label{app:p12}
\begin{table*}[ht]
  \centering
  \caption{First and final instructions for participant P12 in \sone. Content in both instructions is highlighted in blue, and content only in the final instructions is highlighted in yellow.}
  \label{tab:p12_prompts}
  \small        
  \begin{tabular}{@{}p{0.47\textwidth} p{0.47\textwidth}@{}}
    \toprule
    \textbf{First instructions} & \textbf{Final instructions} \\ \midrule
    {\ttfamily
    \colorbox{lights}{\parbox{0.47\textwidth}{Identify whether or not these videos express:}}\newline
    \colorbox{lights}{\parbox{0.47\textwidth}{- One-sided points that solely critique or attack climate change solutions}}\newline
    \colorbox{lights}{\parbox{0.47\textwidth}{- Nihilism about the effectiveness of various climate change solutions, big or small}}\newline
    \colorbox{lights}{\parbox{0.47\textwidth}{- Nihilism about how everyone will die anyway}}\newline
    \colorbox{lights}{\parbox{0.47\textwidth}{- The idea that individual action is unconditionally ineffective as a climate change solution}}\newline
    \colorbox{lights}{\parbox{0.47\textwidth}{- The idea that climate action is too late to be useful}}}
    &
    {\ttfamily
    \colorbox{lights}{\parbox{0.47\textwidth}{Identify whether or not the following videos express:}}\newline
    \colorbox{lights}{\parbox{0.47\textwidth}{- one-sided points that solely critique or attack climate change solutions}}\newline
    \colorbox{lights}{\parbox{0.47\textwidth}{- nihilism about the effectiveness of various climate change solutions, big or small}}\newline
    \colorbox{lights}{- nihilism about how everyone will die anyway}, \colorbox{lightdf}{\parbox{0.47\textwidth}{so climate change is not something to worry or do something about}}\newline
    \colorbox{lights}{\parbox{0.47\textwidth}{- the idea that individual action is unconditionally ineffective as a climate change solution}}\newline
    \colorbox{lights}{\parbox{0.47\textwidth}{- the idea that climate action is being taken too late to be useful}}\newline
    \colorbox{lightdf}{\parbox{0.47\textwidth}{- the idea that humans are too ""dumb"" or ""uneducated"" to change their behavior or solve problems\newline
    - blaming humans unconditionally and without any suggestion that action can still be taken\newline
    - attacking climate activism solely based on any of the traits above\newline
    - how climate change does not fall under the categories of issues that governments or policies can and/or should address\newline
    - negative opinions on climate activists based solely on their appearance or displays of emotion\newline
    - obstructions or obstructionist actors (e.g. climate skeptics) to climate action and/or climate policy and/or climate change perception as legitimate voices to consider in public climate change discourse\newline
    - the view that current climate shifts, patterns, and problems are only ""normal"", ""have always happened"", and/or not actually the issue at hand (e.g. ""We should be focused on climate COOLING instead"") The traits above suggest that the videos may express ""climate pessimism""}}}
    \\ \bottomrule
  \end{tabular}
\end{table*}

\section{Screenshots of Information Features on \sta{} Page}
\label{app:infofeatures}
We include screenshots of all \sta{} page features for additional context. We do not show the \ann{} page features here as they are shown in the main body of the paper.

\begin{figure*}[b]
  \centering
  \includegraphics[width=\textwidth]{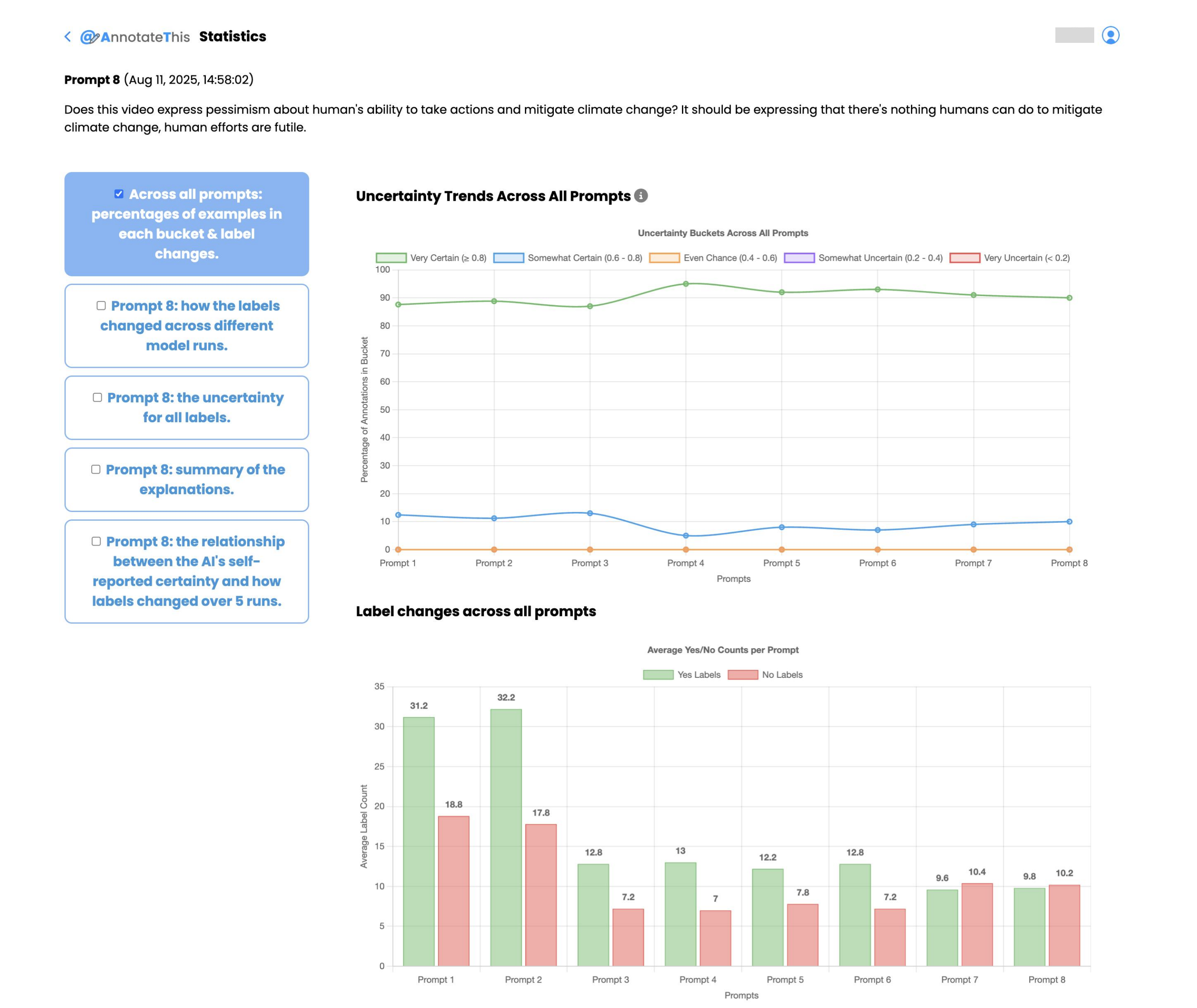}
  \caption{Screenshot of \statren{} feature.}
  \label{fig:screenshot_statren}
\end{figure*}

\begin{figure*}[t]
  \centering
  \includegraphics[width=\linewidth]{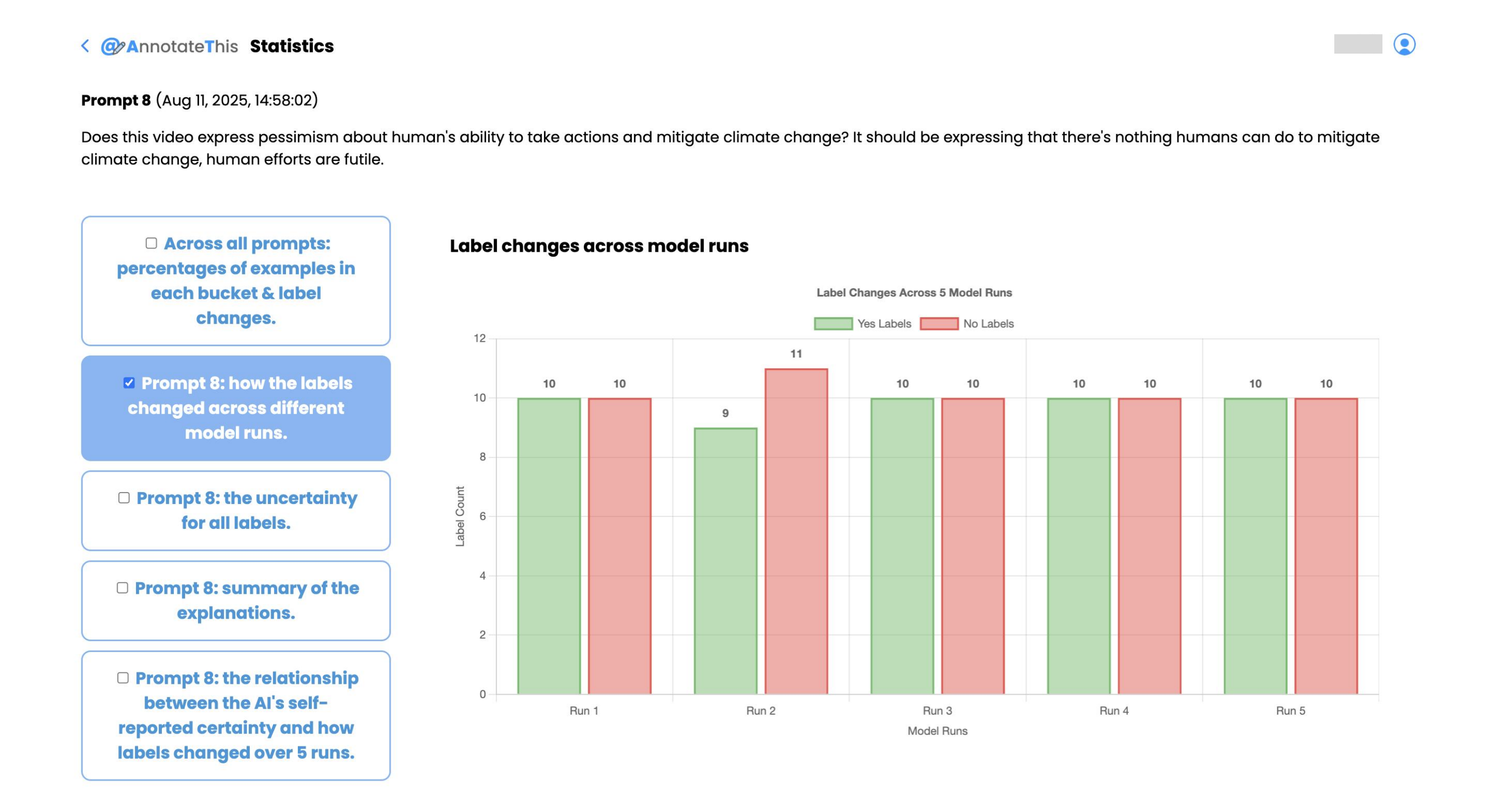}
  \caption{Screenshot of \staruns{} feature.}
  \label{fig:screenshot_staruns}
\end{figure*}

\begin{figure*}[ht]
  \centering
  \includegraphics[width=\linewidth]{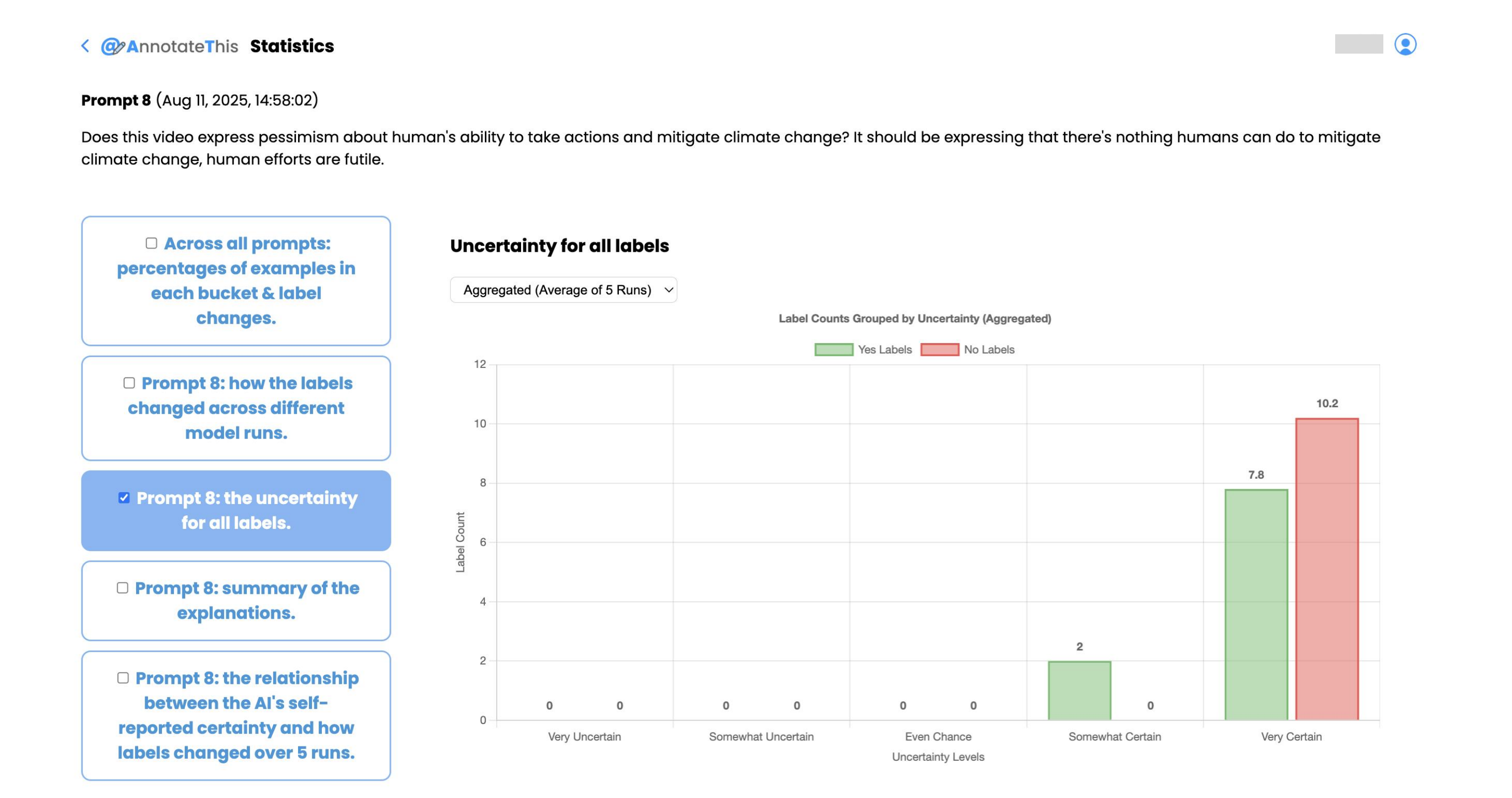}
  \caption{Screenshot of \stabuck{} feature.}
  \label{fig:screenshot_stabucks}
\end{figure*}

\begin{figure*}[ht]
  \centering
  \includegraphics[width=\linewidth]{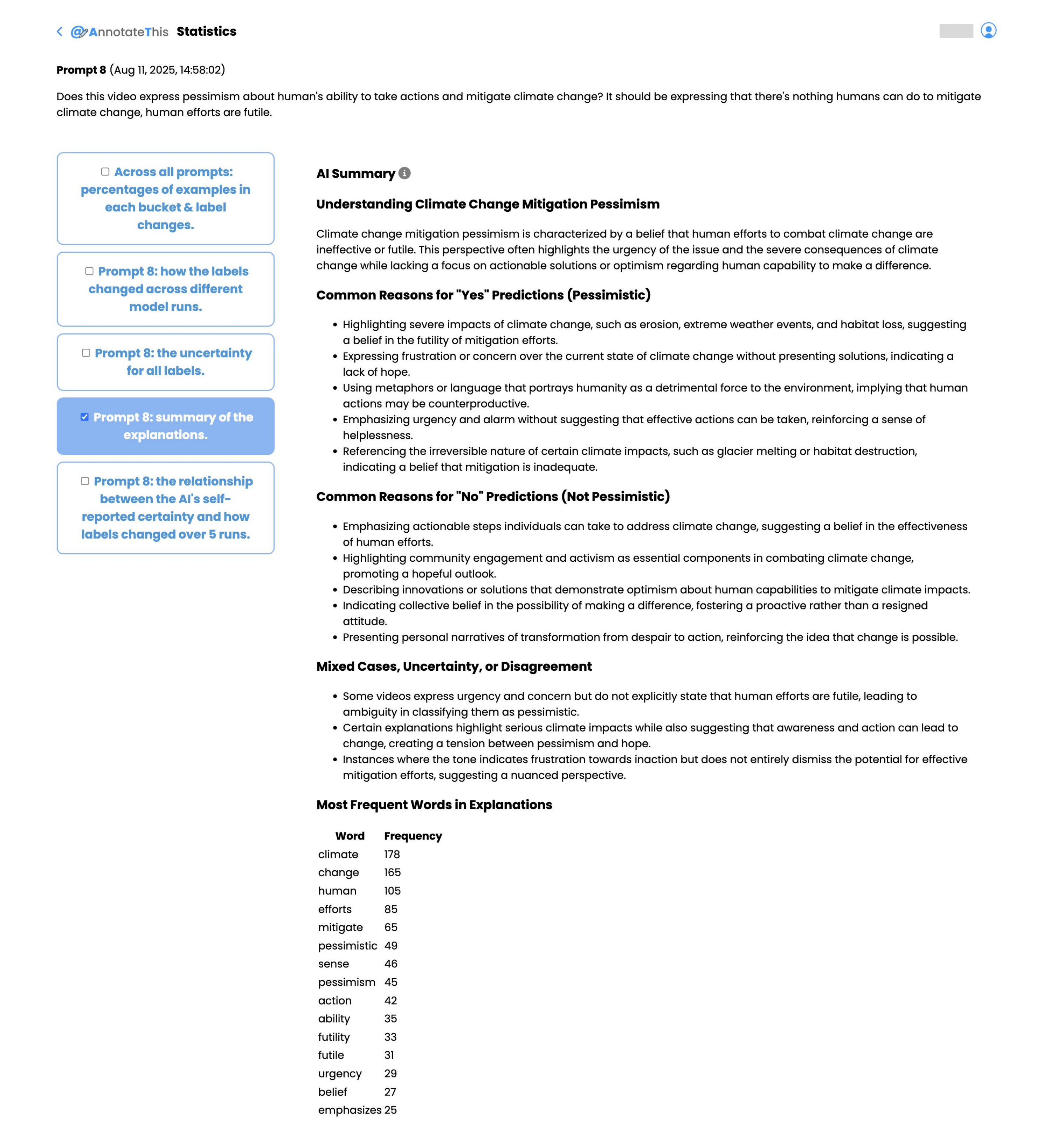}
  \caption{Screenshot of \staexsm{} feature.}
  \label{fig:screenshot_summary}
\end{figure*}

\begin{figure*}[ht]
  \centering
  \includegraphics[width=.7\linewidth]{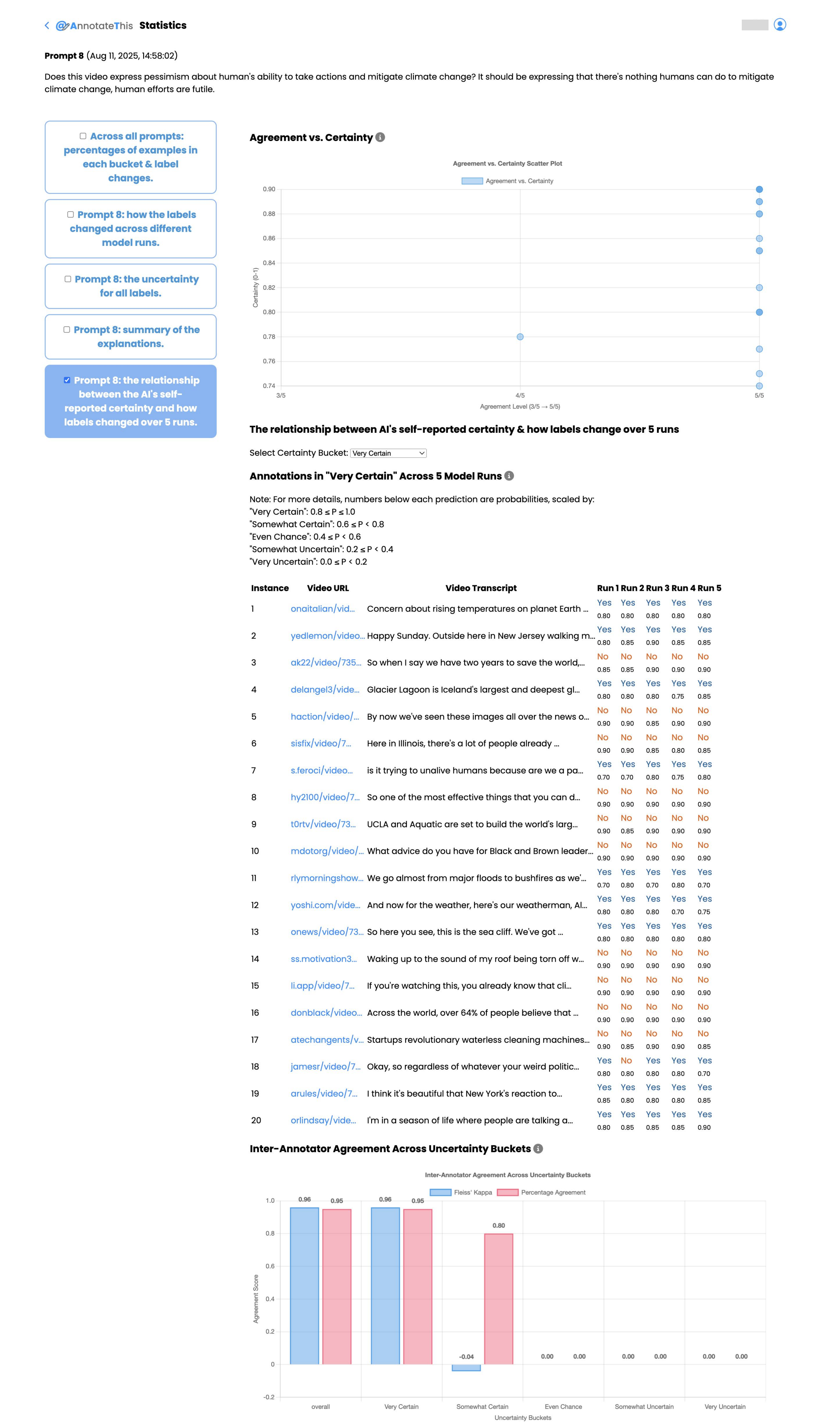}
  \caption{Screenshot of \staagrm{} feature.}
  \label{fig:screenshot_agreement}
\end{figure*}

\begin{description}
    \item[\statren{}] Shown in Fig~\ref{fig:screenshot_statren}
    \item[\staruns{}] Shown in Fig~\ref{fig:screenshot_staruns}
    \item[\stabuck] Shown in Fig~\ref{fig:screenshot_stabucks}
    \item[\staexsm{}] Shown in Fig~\ref{fig:screenshot_summary}
    \item[\staagrm{}] Shown in Fig~\ref{fig:screenshot_agreement}
\end{description}

\section{Additional System Screenshots}
\label{app:system_more}

\begin{figure*}[t]
  \centering
  \includegraphics[width=\linewidth]{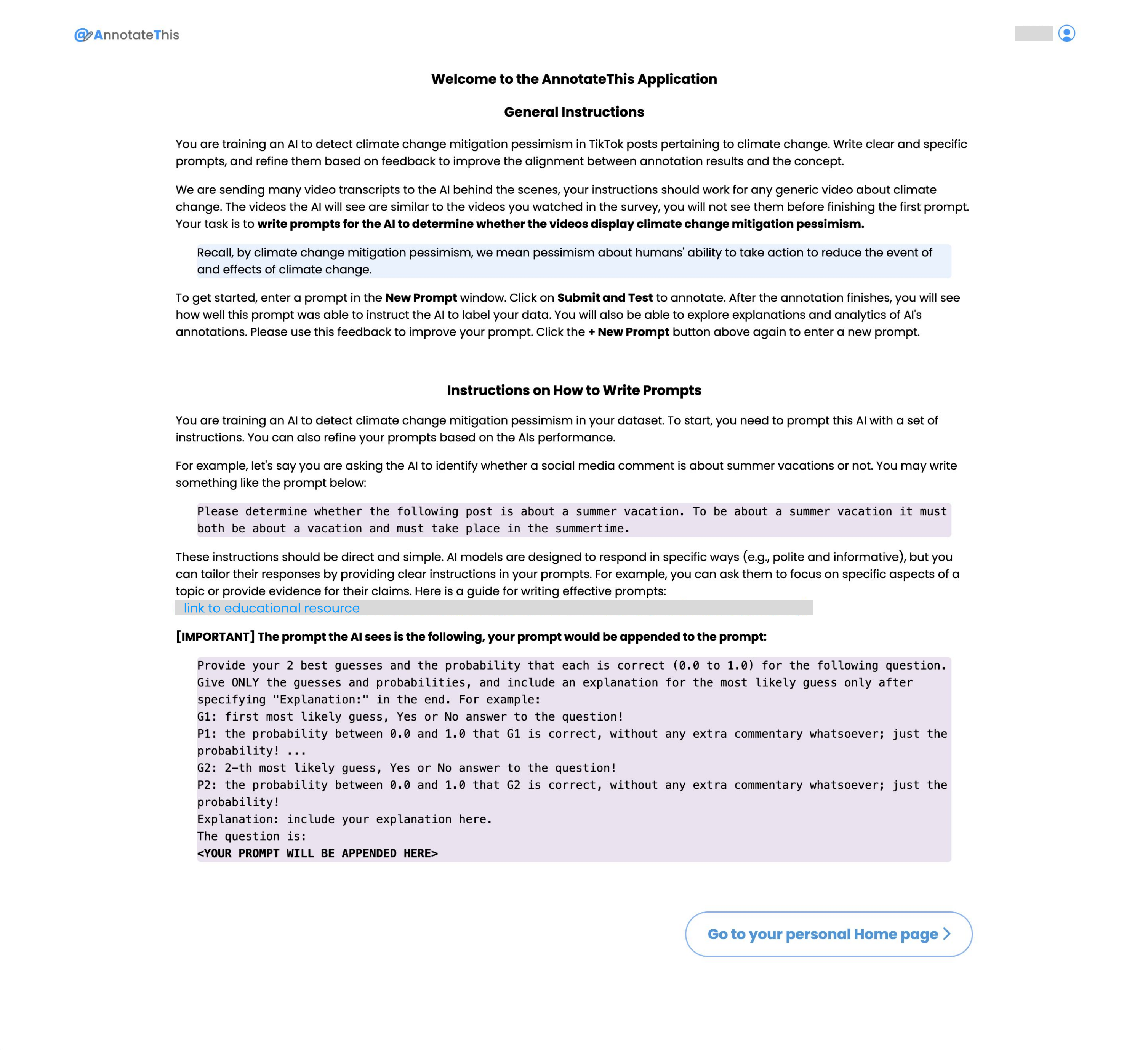}
  \caption{Screenshot of instructions in \ourmethode{}.}
  \label{fig:screenshot_instructions}
\end{figure*}

Here we include screenshots of other pages in \ourmethode{}. Fig~\ref{fig:screenshot_instructions} shows the first page participants see when they log in to the system, which is an instructions page containing general instructions of the task, and instructions on how to write prompts. We also include a link to an external guide on how to write effective prompts, and our system prompt to provide participants with more context and guidance in writing instructions for the LLM.

\begin{figure*}[t]
  \centering
  \includegraphics[width=0.8\linewidth]{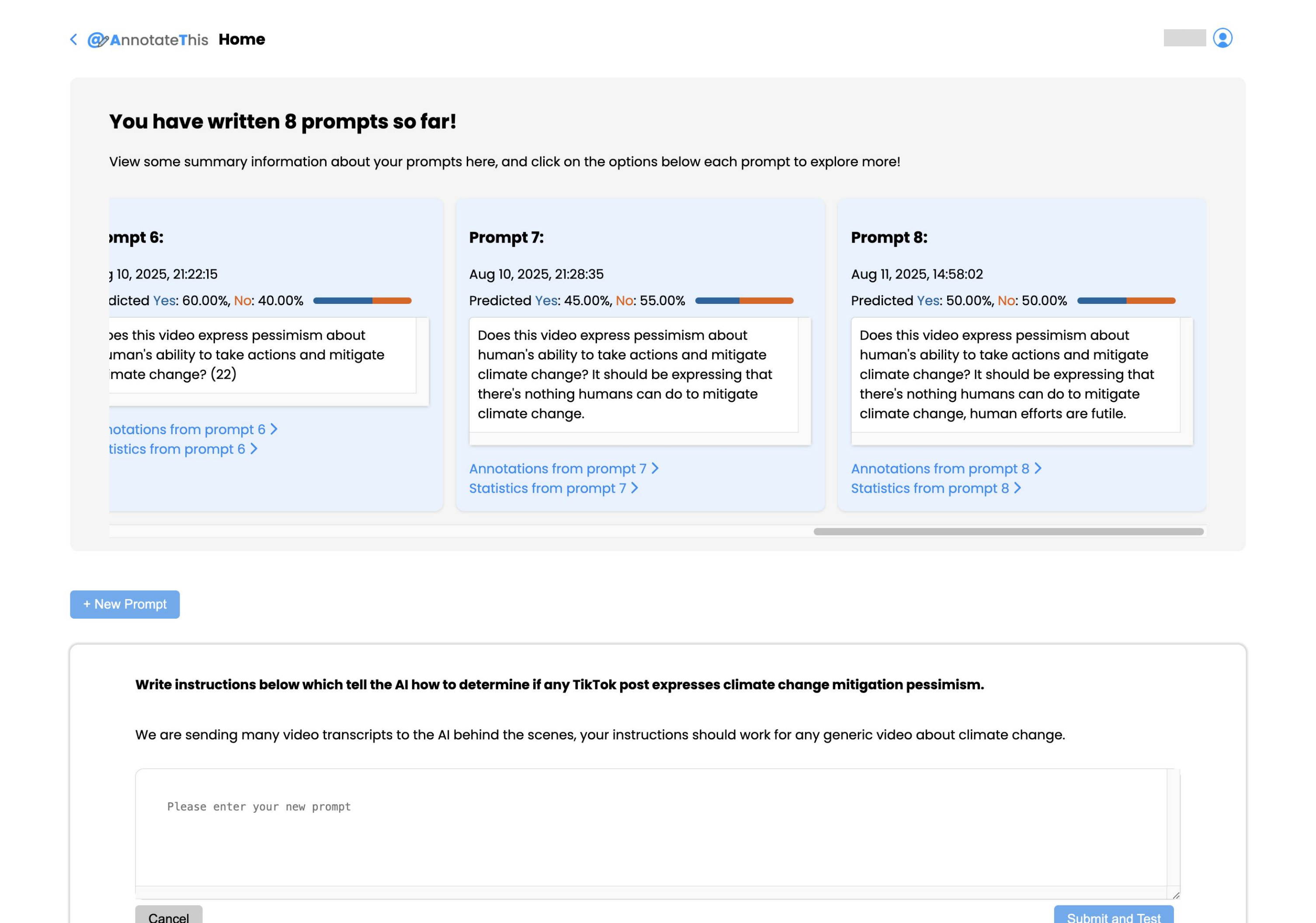}
  \caption{Screenshot of Personal LLM Instruction Hub in \ourmethode{}.}
  \label{fig:screenshot_hub}
\end{figure*}

\begin{description}
    \item[Interface Instructions] Shown in Fig~\ref{fig:screenshot_instructions}
    \item[Personal LLM-I hub] Shown in Fig~\ref{fig:screenshot_hub}
    \item[\anntabl{}] Shown in Fig~\ref{fig:screenshot_stabucks}
\end{description}

\begin{figure*}[t]
  \centering
  \includegraphics[width=0.8\linewidth]{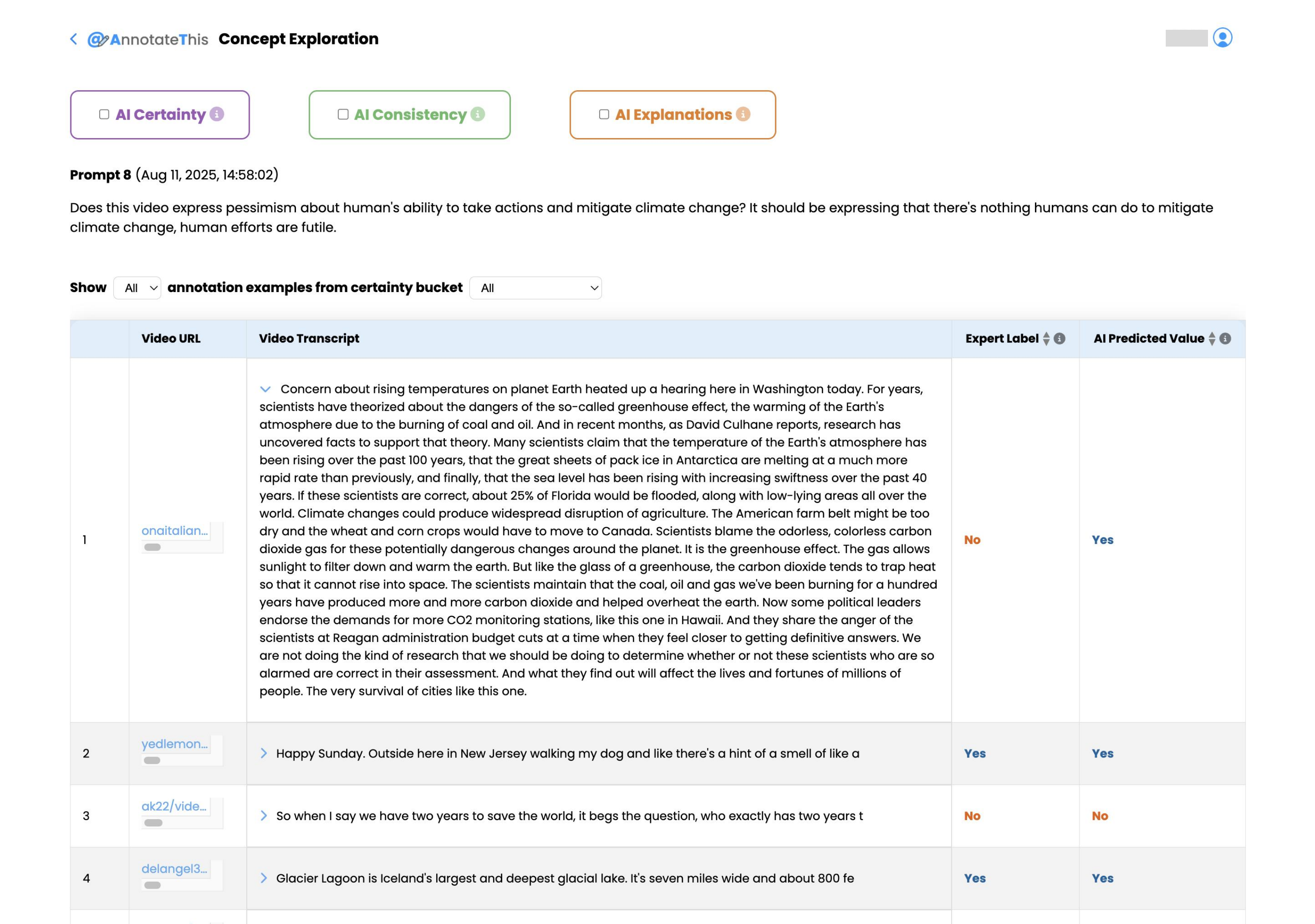}
  \caption{Screenshot of \anntabl{} in \ourmethode{}.}
  \label{fig:screenshot_anntable}
\end{figure*}

\end{document}